\documentclass[lettersize,journal]{IEEEtran}
\usepackage{amsmath,amsfonts}
\usepackage{algorithmic}
\usepackage{algorithm}
\usepackage{array}
\usepackage{textcomp}
\usepackage{stfloats}
\usepackage{url}
\usepackage{verbatim}
\usepackage{graphicx}
\usepackage{cite}
\usepackage{xcolor}
\usepackage{booktabs}
\usepackage{xcolor}
\usepackage{pifont}

\newcommand{\cmark}{\textcolor{green!60!black}{\ding{51}}} 
\newcommand{\xmark}{\textcolor{red!70!black}{\ding{55}}}   

\usepackage{hyperref}
\usepackage{booktabs}       
\usepackage{array}          
\usepackage{enumitem}       
\usepackage{amsmath}        
\usepackage{amsfonts}       
\usepackage{multirow}
\usepackage{subcaption}
\usepackage{amsthm}
\usepackage{amssymb}

\usepackage{array}
\usepackage{pifont}

\newcolumntype{M}[1]{>{\centering\arraybackslash}m{#1}} 

\begin{document}

\title{MU-SHOT-Fi: Self-Supervised  Multi-User Wi-Fi Sensing with Source-free Unsupervised Domain Adaptation}

\author{Ahmed Y. Radwan, {\em Graduate Student Member IEEE},  Hina~Tabassum, {\em Senior Member IEEE}
\thanks{A. Radwan and H. Tabassum are with the department of Electrical Engineering and Computer Science, York University, Toronto, ON M3J 1P3, Canada.
Email: \{ahmedyra, hinat\}@yorku.ca
    }

    }


\maketitle

\begin{abstract}
Deep learning has been widely adopted for WiFi CSI-based human activity recognition (HAR) due to its ability to learn spatio-temporal features in a privacy-preserving and cost-effective manner. However, DL-based models generalize poorly across environments, a challenge that is amplified in multi-user settings where overlapping activities cause CSI entanglement and domain shifts. Moreover, practical deployments often limit access to labeled source data due to privacy constraints, motivating source-free adaptation using only unlabeled target-domain CSI and a pretrained source model.
In this paper, we propose MU-SHOT-Fi, a source-free unsupervised domain adaptation framework for both single-user and multi-user Wi-Fi sensing scenarios. MU-SHOT-Fi employs permutation-invariant set prediction with Hungarian matching during source training, followed by frozen-classifier backbone adaptation in the target domain. To enable stable adaptation under domain shifts without labels, we introduce occupancy-weighted information maximization that prevents model collapse by focusing diversity regularization on likely-occupied slots while excluding the dominant class from marginal entropy. Additionally, we employ binary rotation prediction as spatial self-supervision that exploits CSI frequency-time structure to learn domain-invariant features. For single-user scenarios, we introduce SU-SHOT-Fi by customizing MU-SHOT-Fi through replacing occupancy weighting with standard information maximization and incorporating contrastive predictive coding to exploit temporal consistency. Extensive experiments are conducted on the multi-user WiMANS dataset and single-user Widar 3.0 dataset across cross-environment, cross-frequency, cross-orientation, and combined domain shifts. The results demonstrate that MU-SHOT-Fi effectively recovers multi-user exact-activity classification performance under large domain shifts while maintaining accurate occupancy estimation and preventing collapse toward dominant classes. The source code is publicly available at \url{https://github.com/AhmedRadwan02/mu-shot-fi}.
\end{abstract}

\raggedbottom
\begin{IEEEkeywords}
Wi-Fi sensing, channel state information, source-free domain adaptation, multi-user activity recognition, human activity recognition, permutation-invariant learning, information maximization
\end{IEEEkeywords}

\section{Introduction}

Wi-Fi sensing has emerged as a promising solution for human activity recognition (HAR), offering several advantages over traditional methods such as motion sensors, infrared systems, wearable devices, and camera-based systems. By leveraging existing wireless infrastructure, Wi-Fi sensing enables cost-effective and privacy-preserving sensing without requiring dedicated instrumentation or specialized hardware~\cite{radwan2025tutorial}. Unlike vision- or wearable-based solutions, Wi-Fi sensing captures channel state information (CSI), which reflects variations in wireless signals as they interact with people and surroundin objects~\cite{ding2024multiple}. This enables applications including HAR, localization, gesture recognition, and vital sign monitoring~\cite{shi2022environment,li2021deep,raja2018wibot,shirakami2021heart}. Additionally, Wi-Fi sensing can operate in non-line-of-sight (NLOS) conditions, giving it a distinct advantage over camera-based systems that require direct line of sight (LOS). This capability to sense through walls and obstacles makes it particularly versatile for indoor applications.

Prior work in Wi-Fi sensing relied on traditional signal processing methods such as Fresnel Zone modeling~\cite{wu2022wifi}, Angle of Arrival (AoA), and Time of Flight (ToF) to characterize amplitude attenuation and phase shifts in multipath channels~\cite{ahmed2018estimating}. However, these approaches depend heavily on hand-crafted statistical features to represent signal variations and human activities, which becomes infeasible when handling complex movements in practice~\cite{yousefi2017survey}. Furthermore, traditional signal processing methods typically treat each CSI sample independently, failing to capture temporal dependencies in the signal. These limitations hinder generalization across different environments~\cite{brunello2025time}, often necessitating environment-specific calibration for each new deployment~\cite{cominelli2023exposing}.

Recently, deep learning (DL) has significantly improved the quality of CSI-based Wi-Fi sensing by learning feature representations directly from raw data without requiring manual feature engineering~\cite{yousefi2017survey,ma2019wifisensing}. DL models can capture both spatial patterns across CSI subcarriers and antennas, as well as temporal dependencies in signal sequences, enabling recognition of complex activities that unfold over time~\cite{chen2018wifi}. 
{This feature learning allows DL approaches to handle complex movement patterns and environment, without the need for handcrafted calibration compared to traditional methods.

Despite significant progress in DL-based Wi-Fi sensing, a critical gap remains between controlled laboratory demonstrations and practical deployments~\cite{ma2019wifisensing, tan2022commodity}. Most existing research works remain limited to  \textit{single-user} sensing environments~\cite{yousefi2017survey}, real-world applications such as smart homes, office buildings, and healthcare facilities inherently involve multiple users that perform activities concurrently. When multiple users perform activities jointly, their movements create overlapping  CSI patterns~\cite{huang2024wimans,tan2019multitrack}. This signal entanglement makes it difficult to attribute observed CSI variations to individual users. Existing multi-user sensing approaches address this through signal decomposition~\cite{wang2017tensorbeat}, auxiliary task coupling, such as user identity prediction~\cite{huang2024wimans} or location estimation~\cite{rizk2025multisensex}, or specialized hardware~\cite{tan2019multitrack}. However, these methods often rely on additional annotations and make restricted assumptions about signal separability.

Beyond the \textit{multi-user challenge}, existing DL-based Wi-Fi sensing solutions suffer from \textit{generalization} issues when the target domain shifts~\cite{strohmayer2024data,brunello2025time}. Models trained in specific environments often experience significant accuracy degradation when deployed in unseen environments. These domain shifts arise from changes in the physical environment (furniture layout, room structure), user characteristics (different people, body types, movement styles), hardware differences (device types, antenna configurations), and temporal factors (signal drift, environmental conditions)~\cite{cominelli2023exposing}. The inherent sensitivity of CSI signals to environmental conditions means that even minor changes can cause substantial distribution shifts~\cite{strohmayer2024data}. 

Importantly, \textit{domain adaptation} becomes more challenging in \textit{multi-user wireless sensing} scenarios due to dominant distribution shifts caused by multi-user CSI signals, and addressing this challenge is the focus of this paper.

Recently, a couple of research works proposed unsupervised domain adaptation (UDA)  strategies where the models are trained using labeled source domain data alongside unlabeled target domain data \cite{zhang2023unsupervised, chen2022fidora, jiao2025robust}. 
However, UDA strategies exhibit fundamental limitations in wireless sensing scenarios as they require access to labeled source domain data, which is often infeasible in practice due to users' privacy constraints~\cite{liang2020we} and the need of online CSI annotations in wireless sensing. 

{To address the limitations of UDA, recently source-free unsupervised domain adaptation (SFUDA) solutions are gaining attention. SFUDA enables adapting a model trained in source domain to target domain using only unlabeled target-domain data, without requiring access to source-domain samples or labels~\cite{li2024comprehensive}. To our knowledge, Wi-SFDAGR~\cite{yan2025wi} is the only  SFUDA solution proposed very recently for single-user gesture recognition using unsupervised clustering, relying on the assumption that samples with similar features should have similar labels. That is, each CSI sample corresponds to one person performing one activity, thus all "push" gestures cluster together because they produce similar CSI signal patterns. 

Nevertheless, an assumption that is reasonable in \emph{single-user} sensing does not directly extend to \emph{multi-user} sensing.
In the single-user case, each sample is associated with a single activity label $y \in \mathcal{A}$, where $\mathcal{A}$ is the set of $K$ possible activities, so nearby feature representations are expected to share the same label. In contrast, in the multi-user case, each CSI sample is annotated by a vector of joint activities of up to $M$ users $\tilde{\mathbf{y}} = [y_1, \ldots, \emptyset, \ldots, y_M]$, where 
$M$ is the maximum occupancy (e.g., $M{=}6$ in our experiments) and $\emptyset$ indicates an unoccupied slot when fewer than $M$ users are present.

Because users are not indexed in a canonical way, $\mathbf{y}$ is only defined \emph{up to permutation} of its entries; different orderings can describe the same underlying multi-user state.
For example, $[\texttt{walk},\texttt{jump},\emptyset]$ and $[\texttt{jump},\texttt{walk},\emptyset]$ are equivalent.
Thus, two samples with very similar features may correspond to different \emph{ordered} label vectors even when the underlying set of activities is identical, which complicates the application of clustering-based adaptation as in Wi-SFDAGR~\cite{yan2025wi}.

To address the aforementioned challenges, our contributions can be summarized as follows:
\begin{enumerate}

\item We present \textsc{MU-SHOT-Fi} (\textbf{M}ulti-\textbf{U}ser \textbf{S}ource \textbf{H}ypothesis \textbf{O}ptimization via \textbf{T}ransfer for \textbf{Wi-Fi} sensing), an SFUDA framework for both single-user and multi-user sensing using  CSI amplitude or phase ratio. Unlike~\cite{shi2020towards,yan2025wi,zhang2023unsupervised},  our approach is agnostic to preprocessing of CSI data.

\item We introduce occupancy-weighted information maximization to enable stable adaptation in multi-user sensing scenarios and to mitigate the class imbalance caused by the unoccupied (''no person'') slots, which dominate many datasets due to the scarcity of samples at higher user counts. Unlike standard diversity regularization that treats all output slots uniformly, our objective weights each slot according to its estimated occupancy probability and excludes the \texttt{no person} class from the marginal entropy term. This design prevents the model from collapsing toward dominant no-person-class predictions and promotes balanced adaptation across active users.

\item {\textsc{MU-SHOT-Fi} employs (1) {permutation-invariant 
set-based prediction across $M$ slots during source domain 
training, building on set prediction principles 
from~\cite{carion2020end} and its application to 
multi-user Wi-Fi sensing in~\cite{mohammadi2026amar}}}, and (2) target adaptation with frozen classifier while updating the backbone through rotation-based spatial self-supervision that exploits CSI's frequency-time structure to learn domain-invariant features, combined with occupancy-weighted diversity regularization.

\item We customize \textsc{MU-SHOT-Fi} for single-user sensing by employing standard information maximization instead of occupancy weighted information maximization, activating k-nearest centroid pseudo-labeling which is suitable for single-user CSI data with no class imbalance, and adding Contrastive Predictive Coding (CPC) loss term that exploits temporal consistency in the adaptation stage.

\item We evaluate the proposed framework on two datasets, i.e., WiMANS~\cite{huang2024wimans} across cross-environment, cross-frequency, and combined shifts, and on Widar 3.0~\cite{zhang2021widar3} across cross-room, cross-torso, and cross-face settings.
On WiMANS combined shifts, \textsc{MU-SHOT-Fi} achieves 41.97\% slot-wise accuracy and 2.12 occupancy MAE compared to 19.61\% and 2.48 for source-only models.
On Widar 3.0, it improves single-user accuracy from 80.18\% to 85.75\% averaged across settings.
We further analyze limitations of permutation-invariant architectures under unsupervised adaptation.
\end{enumerate}

The remainder of this paper is organized as follows. Section~\ref{sec:background} provides fundamentals of  CSI and reviews existing work on domain adaptation techniques. Section~\ref{sec:method} presents the problem formulation and introduces the proposed framework. Section~\ref{subsec:multiuser} focuses on describing MU-SHOT-Fi and Section~\ref{subsec:singleuser}  customizes MU-SHOT-Fi for single-user scenarios. Section~\ref{sec:experiments} describes experimental setup. Section~\ref{sec:results} depicts numerical results comparing model performance, followed by ablation studies analyzing the impact of key design choices. Finally, Section~\ref{sec:conclusion} concludes the paper.

\section{Preliminaries and Related Work}
\label{sec:background}
In this section, we describe CSI fundamentals and domain shift in Wi-Fi sensing, then review key existing UDA and SFUDA approaches that are also summarized in Table~\ref{tab:related_work_summary}.

\subsection{Domain Shift in CSI-Based Wi-Fi Sensing}
\label{subsec:wifi_csi}
Wi-Fi sensing exploits CSI from commodity devices for privacy-preserving HAR. In a MIMO-OFDM system with $N_t$ transmit and $N_r$ receive antennas over $N_{sc}$ subcarriers, the received signal on subcarrier $k$ at time $t$ is $\mathbf{y}_{k}(t) = \mathbf{H}_{k}(t)\,\mathbf{s}_{k}(t) + \boldsymbol{\eta}_{k}(t)$, where $\mathbf{s}_{k}(t)\in\mathbb{C}^{N_t}$ is the transmitted symbol vector, $\boldsymbol{\eta}_{k}(t)\in\mathbb{C}^{N_r}$ denotes additive noise, and $\mathbf{H}_{k}(t) \in \mathbb{C}^{N_r\times N_t}$ is the CSI matrix. Let $H_{m,n}(k,t)$ denotes the CSI from transmit antenna $n\in\{1,\dots,N_t\}$ to receive antenna $m\in\{1,\dots,N_r\}$ on subcarrier $k$ at time $t$, so that $\mathbf{H}_{k}(t)=[H_{m,n}(k,t)]_{m=1,n=1}^{N_r,N_t}$. Each CSI entry representing the complex channel frequency response (CFR) is $H_{m,n}(k,t) = \lvert H_{m,n}(k,t) \rvert\, e^{j \phi_{m,n}(k,t)}$, where $\phi_{m,n}(k,t)\triangleq \angle H_{m,n}(k,t)$. Commodity NICs expose CSI as tensors, i.e.,
\begin{equation}
\label{eq:x_raw}
\mathbf{x}_{\text{raw}} \triangleq [H_{m,n}(k, t)]_{t,m,n,k} \in \mathbb{C}^{T \times N_r \times N_t \times N_{sc}}, 
\end{equation}
across time, antennas, and subcarriers~\cite{luo2018channel}. We use $\mathbf{x} \in \mathbb{R}^{1 \times F \times T}$ to denote its preprocessed representation (e.g., amplitude or phase extraction with spatial flattening, detailed in Section~\ref{sec:method}), where $F = N_{sc} \cdot N_r \cdot N_t$ is the flattened  dimension.

Indoor propagation is dominated by multipath effects. A common model that expresses the CFR as a superposition of $L$ propagation paths is given below:
\begin{equation}
H_{m,n}(k,t) = \sum_{\ell=1}^{L} \alpha_{\ell,mn}(t)\, e^{-j2\pi f_k \tau_{\ell,mn}(t)}\, e^{j2\pi f_{D,\ell} t},
\label{eq:multipath_model}
\end{equation}
where $f_k$ is the center frequency of subcarrier $k$, subscript $mn$ denotes the transmit--receive antenna pair $(m,n)$, $\alpha_{\ell,mn}(t)\in\mathbb{C}$ and $\tau_{\ell,mn}(t)$ are the time-varying complex attenuation and delay of path $\ell$, respectively, and $f_{D,\ell}$ (Hz) is the Doppler shift producing the time-varying phase term $e^{j2\pi f_{D,\ell} t}$. Small changes in path delay induce large phase rotations, i.e., a delay perturbation $\Delta\tau$ yields
\begin{equation}
\Delta\phi_k \approx -2\pi f_k \Delta\tau \quad (\mathrm{mod}\ 2\pi).
\label{eq:phase_delay_compact}
\end{equation}
 Consequently, modest changes in room layout, materials, user placement, carrier frequency, or device calibration substantially alter CSI statistics, causing abrupt domain shifts in both amplitude and phase~\cite{wei2025survey,radwan2025tutorial,hasanzadeh2024enhancing}.

In multi-user scenarios, Eq.~\eqref{eq:multipath_model} aggregates contributions from all active users simultaneously. Unlike single-user settings where the $L$ paths primarily reflect one person's movements, multi-user environments superpose path sets from each individual, where path parameters $\{\alpha_{\ell,mn}(t), \tau_{\ell,mn}(t), f_{D,\ell}\}$ reflect the combined influence of multiple people performing different activities~\cite{tan2019multitrack}. This signal entanglement makes it fundamentally difficult to attribute observed CSI variations to specific users, as one person's movements directly affect how another person's activities appear in the wireless channel~\cite{huang2024wimans}.

Domain shift compounds these multi-user challenges by altering both individual activity signatures and their interactions. Consider the cross-frequency scenario in WiMANS where models trained at 2.4~GHz must generalize to 5~GHz. The wavelength changes from 12.5~cm to 5~cm, fundamentally altering how overlapping user activities interfere in the multipath channel. From Eq.~\eqref{eq:phase_delay_compact}, phase rotations scale directly with carrier frequency: a given path delay $\Delta\tau$ induces phase shifts $\Delta\phi_{5\text{GHz}} / \Delta\phi_{2.4\text{GHz}} = 5.0/2.4 \approx 2.08$ times larger at the higher frequency. This frequency-dependent phase sensitivity means that entangled multi-user patterns exhibit domain shifts that cannot be addressed by single-user  adaptation techniques~\cite{wang2017tensorbeat}.

\begin{table*}[t]
\centering
\caption{Summary of Existing Works on Domain Adaptation for Wi-Fi Sensing}
\label{tab:related_work_summary}
\small
\renewcommand{\arraystretch}{0.9}
\setlength{\tabcolsep}{3pt}
\begin{tabular}{
p{2.5cm}  
p{2.8cm}  
p{2.8cm}  
p{1.8cm}  
M{1.3cm}  
M{0.9cm}  
M{1.2cm}  
M{1.2cm}  
M{2.0cm}  
}
\toprule
\textbf{Reference} & \textbf{Task} & \textbf{Environment} & \textbf{Input} &
\textbf{Multi-User} &
\textbf{SSL at Target} & \textbf{Agnostic Source Model} & \textbf{Agnostic Preprocessing} & \textbf{Open Access Data} \\
\midrule
\multicolumn{9}{l}{\textit{Unsupervised Domain Adaptation (UDA)}}\\
\midrule
\cite{zhang2023unsupervised} & Gesture Recognition & Lab (Classroom, Hall, Office) & Phase Ratio (BVP)
& \xmark & \xmark & \cmark & \cmark & Widar 3.0 \\
\cmidrule(lr){2-9}
\cite{shi2020towards} & User Authentication \& HAR & Residential Apartment \& Office & Amplitude (Time Domain\& Spectrogram)
& \xmark & \xmark & \cmark & \xmark & \xmark \\
\cmidrule(lr){2-9}

WiAi-D~\cite{liang2023wiai} & Person Recognition & Indoor Floor & Amplitude
& \xmark & \xmark & \xmark & \xmark & \xmark \\
\cmidrule(lr){2-9}

Fidora~\cite{chen2022fidora} & Localization & Office & Amplitude
& \xmark & \xmark & \xmark & \xmark & \xmark \\
\cmidrule(lr){2-9}

DF-Loc~\cite{jiao2025robust} & Localization & Classroom \& Office & Amplitude \& Phase
& \xmark & \xmark & \xmark & \xmark & \xmark \\
\midrule

\multicolumn{9}{l}{\textit{Source-Free Unsupervised Domain Adaptation (SFUDA)}}\\
\midrule
Wi-SFDAGR~\cite{yan2025wi} & Wi-Fi Gesture Recognition & Lab (Classroom, Hall, Office) & Phase Ratio
& \xmark & \xmark & \cmark & \xmark & XRF55, Widar 3.0 \\
\cmidrule(lr){2-9}

\textbf{MU-SHOT-Fi} & \textbf{HAR \& Gesture Recognition} & \textbf{Lab (Classroom, Hall, Meeting Room)} & \textbf{Amplitude, Phase Ratio}
& \cmark & \cmark & \cmark & \cmark &  WiMANS, Widar 3.0 \\
\bottomrule
\end{tabular}
\end{table*}

\subsection{Related Work in Wi-Fi Sensing}
To date, a handful of research works addressed domain generalization in single-user Wi-Fi sensing using UDA techniques. UDA employs labeled source data and unlabeled target data for training, reducing source--target discrepancy while preserving task-discriminative structure. Most UDA solutions follow a ``learn-and-align'' paradigm where a backbone is trained with labeled source data while the loss function encourages target features to match the source distribution.

Recently, \cite{shi2020towards} proposed an environment-independent user authentication system using CSI amplitude and spectrograms processed by CNN-based feature extractors. They train two classifiers (user identity and HAR) with adversarial domain adaptation: a domain discriminator predicts the data domain from learned features while the feature extractor is simultaneously trained to prevent discriminator's success. This minimax objective retains identity- and activity-relevant information while suppressing environment-specific cues. Similarly, WiAi-ID~\cite{liang2023wiai} adopted adversarial domain adaptation for passive person identification, making identity features robust to appearance changes (clothing, carried items) rather than environmental changes, using CSI amplitude from single-person walking traces. Both methods evaluate on single-user scenarios.

Adversarial domain adaptation is most effective when domain shifts alter irrelevant signal characteristics while label-relevant features remain stable across domains. In multi-user HAR, however, CSI reflects an entangled superposition of multiple users' motions, and environment- or frequency-dependent variations are tightly coupled with the cues required for user separation~\cite{strohmayer2024data,cominelli2023exposing}. Enforcing domain invariance in this setting can suppress information essential for disentangling users, leading to negative transfer\footnote{Negative transfer occurs when adaptation degrades performance: the adapted model performs worse than a non-adapted baseline (e.g., source-only) because the adaptation objective suppresses label-relevant information along with domain-specific factors.}. In contrast, single-user settings better satisfy the assumptions of adversarial adaptation, as identity- and activity-relevant features can be isolated.

Beyond adversarial alignment, reconstruction-based objectives have been explored for UDA. Fidora~\cite{chen2022fidora} addressed indoor localization by classifying which of eight predefined locations a person occupies based on CSI fingerprints, using a variational autoencoder (VAE) for CSI data augmentation and a joint classification-reconstruction network for domain adaptation across environmental changes. However, reconstruction methods assume stable, deterministic mappings between CSI patterns and spatial/activity representations. Multi-user scenarios violate this assumption through signal entanglement~\cite{tan2019multitrack}-the same CSI pattern can correspond to different user-activity configurations depending on the number of users, their relative positions, and movement synchronization.

DF-Loc~\cite{jiao2025robust} improved localization accuracy with Multi-Source UDA (MUDA), leveraging multiple environments as source domains. Using both CSI amplitude and phase, DF-Loc employed two-stage alignment: first aligning feature distributions, then aligning regressor outputs for prediction consistency, plus adversarial learning with domain discriminators to enhance domain-invariant feature extraction. However, multi-source transfer risks negative transfer when sources poorly match the target. In multi-user scenarios, variable occupancy across source domains introduces biased activity distributions that degrade adaptation performance.

Lastly,~\cite{zhang2023unsupervised} presented one of the first UDA methods for RF-based gesture recognition on the Widar 3.0 dataset. They extracted body-coordinate velocity profiles (BVP) from CSI phase ratios and employed (1) pseudo-labeling to generate pseudo-labels for unlabeled target data, enabling cross-entropy training, and (2) consistency regularization that enforces prediction consistency between original BVP features and augmented versions. However, pseudo-labeling can lead to class imbalance bias-converging toward dominant classes when the dataset is skewed. This is also a challenge for multi-user scenarios, where variable user counts create imbalanced activity distributions where empty-room or single-user samples may dominate, causing the model to under-represent occupancy states and degrade performance on classes with less samples.

Overall, existing UDA solutions focus exclusively on single-user cases and assume access to source domain data, raising privacy concerns when data is confidential. Moreover, adversarial learning  while effective for single-user alignment-does not guarantee applicability in multi-user settings due to signal entanglement that violates domain discriminator assumptions.

More recently, SFUDA has emerged as an alternative that adapts pretrained source models using only unlabeled target data~\cite{liang2020we}. SFUDA methods typically employ entropy minimization, pseudo-labeling, or neighborhood consistency~\cite{liang2020we,liang2021source}. However, applying these to multi-user Wi-Fi sensing introduces three fundamental challenges: (1) class imbalance: standard information maximization treats all output dimensions uniformly, but ``no person'' slots dominate when users are less than $M$~\cite{huang2024wimans}, causing collapse toward the dominant class; (2) permutation invariance: clustering-based pseudo-labeling cannot specify which activity belongs to which slot, while Hungarian matching requires explicit slot-level targets unavailable during UDA; (3) signal entanglement: neighborhood consistency assumes stable feature-to-label mappings, but multi-user CSI reflects superposed activities where identical features may correspond to different user configurations~\cite{tan2019multitrack}.

Wi-SFDAGR~\cite{yan2025wi} addresses single-user gesture recognition through clustering-based adaptation with an attraction--dispersion objective and uncertainty-weighted neighbor selection based on prediction entropy. While effective for single-user settings with balanced gesture distributions, this approach assumes (i) samples with similar features belong to the same class, enabling direct cluster-to-label assignment, and (ii) each sample maps to a single activity label. Both assumptions break down in multi-user scenarios due to class imbalance and permutation invariance, i.e., similar features can correspond to different slot orderings of the same underlying activity set (e.g., [walk, sit, $\emptyset$] and [sit, walk, $\emptyset$] are equivalent but produce different feature-label pairs).

Overall, existing UDA and SFUDA methods primarily target single-user classification under moderate domain gaps~\cite{fang2024source,li2024comprehensive}. Multi-user sensing suffers from: (i) class imbalance from unoccupied slots, (ii) signal entanglement invalidating stable feature mappings, and (iii) variable occupancy requiring structured set predictions rather than single-label outputs.

\begin{figure*}[!t]
    \centering
    \includegraphics[width=\linewidth]{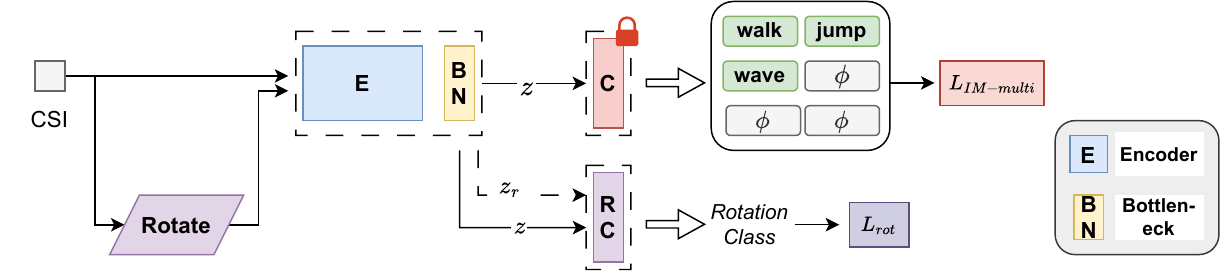}
    \caption{Architecture of the proposed MU-SHOT-Fi source-free unsupervised domain adaptation framework for multi-user Wi-Fi HAR. During source training, permutation-invariant set prediction with Hungarian matching handles variable occupancy across $M$ slots. During target adaptation, the classifier is frozen while the backbone is updated via (i) occupancy-weighted information maximization to prevent collapse toward the dominant no-person class, and (ii) rotation-based spatial self-supervision exploiting CSI frequency-time structure for domain-invariant feature learning.}
    \label{fig:mu_shoft_fi}
\end{figure*}

\section{MU-SHOT-Fi: Set-based Multi-User HAR  Formulation and Source Training}
\label{sec:method}

The proposed \textsc{MU-SHOT-Fi} (Fig.~\ref{fig:mu_shoft_fi})  addresses multi-user sensing through two key components: (1) permutation-invariant set-based prediction across $M$ slots during source training via Hungarian matching (Section~\ref{subsec:hungarian}), and (2) target adaptation with frozen classifier while updating the backbone through rotation-based spatial self-supervision  (Section~\ref{subsec:multiuser}).  Specifically, this section first details the preprocessing of the CSI data, then presents the set-based HAR problem formulation, and finally describes the source-domain training mechanism for multi-user sensing.

\subsection{Data Preprocessing}
As established in Section~\ref{subsec:wifi_csi}, multi-user Wi-Fi sensing presents fundamental challenges due to signal entanglement, where the multipath model (Eq.~\ref{eq:multipath_model}) aggregates contributions from all active users simultaneously, making it difficult to attribute observed CSI variations to specific individuals~\cite{tan2019multitrack,huang2024wimans}. 
Unlike single-user settings, multi-user scenarios involve predicting an unordered activity set $\mathbf{y} = \{y_1, \ldots, y_{N_p}\}$, where $y_i \in \mathcal{A}$ represents an activity from $K$ possible classes and $N_p$ denotes the true number of people present. Critically, $N_p$ is unknown and varies across samples, creating a variable-size prediction problem. This inherent permutation invariance---where one person walking while another sits has the same semantic meaning as one person sitting while another walks---motivates formulating the problem as set prediction rather than ordered sequence prediction.

From commodity Wi-Fi devices, we obtain raw CSI measurements as complex-valued tensors $\mathbf{x}_{\text{raw}} \in \mathbb{C}^{T \times N_r \times N_t \times N_{sc}}$ (Eq.~\ref{eq:x_raw}), representing temporal samples, receive antennas, transmit antennas, and subcarriers. We reshape the spatial dimensions, yielding the input representation $\mathbf{x} \in \mathbb{R}^{1 \times F \times T}$, where $F = N_{sc} \cdot N_r \cdot N_t$ is the flattened spatial dimension and $T$ is the temporal length. 
To handle variable occupancy, we pad the ground-truth activity set $\mathbf{y}$ to a fixed-size vector $\tilde{\mathbf{y}} \in (\mathcal{A} \cup \{\emptyset\})^M$ of length $M$ using the "no person" token: $\tilde{\mathbf{y}} = [y_1, \ldots, y_{N_p}, \emptyset, \ldots, \emptyset]$, where $\tilde{\mathbf{y}}$ contains the $N_p$ actual activities and $M-N_p$ instances of $\emptyset$ (the ordering is considered up to permutation, since training uses permutation-invariant bipartite matching). This allows the model to predict $\emptyset$ for unoccupied slots, reducing the hypothesis space from $(K)^{N_p}$ ordered sequences to $\binom{K+N_p-1}{N_p}$ unordered sets{~\cite{mohammadi2026amar}}.

\subsection{Source Model and Set-Based Multi-User HAR Formulation}
\label{subsec:set_prediction}

We consider SFUDA for CSI-based multi-user sensing, where a model trained on a labeled source domain must adapt to an unlabeled target domain without access to the source dataset. Formally, let $\mathcal{D}_s=\{(\mathbf{x}_i^s,\mathbf{y}_i^s)\}_{i=1}^{n_s}$ denotes $n_s$ labeled source samples drawn from distribution $P_s(\mathbf{X},\mathbf{Y})$, and let $\mathcal{D}_t=\{\mathbf{x}_j^t\}_{j=1}^{n_t}$ denotes $n_t$ unlabeled target samples drawn from marginal distribution $P_t(\mathbf{X})$. The source and target domains exhibit distribution shift, i.e., $P_t(\mathbf{X},\mathbf{Y})\neq P_s(\mathbf{X},\mathbf{Y})$. During adaptation, only the pre-trained source model and target data $\mathcal{D}_t$ are available; the source data $\mathcal{D}_s$ is inaccessible.

The source model consists of three components parameterized by $\theta_s = \{\eta_s, \psi_s, \phi_s\}$, i.e., 
a feature extractor $F_{\eta_s}: \mathbb{R}^{1 \times F \times T} \to \mathbb{R}^{d_f}$, a bottleneck (linear) layer $B_{\psi_s}: \mathbb{R}^{d_f} \to \mathbb{R}^{d_b}$, and a classifier $C_{\phi_s}: \mathbb{R}^{d_b} \to \mathbb{R}^{K}$ (single-user) or $C_{\phi_s}: \mathbb{R}^{d_b} \to \mathbb{R}^{M \times (K+1)}$ (multi-user). 
The complete forward pass is $f_{\theta_s}(\mathbf{x}) = C_{\phi_s}(B_{\psi_s}(F_{\eta_s}(\mathbf{x})))$. For single-user tasks, the classifier outputs $K$ activity class logits. For multi-user sensing, it outputs $M$ slots with $K{+}1$ classes each ($K$ activities plus ``no person''), where $M$ is the maximum occupancy.

We then formulate multi-user activity recognition as a set 
prediction problem, inspired by~\cite{carion2020end} {and 
its application to multi-user Wi-Fi sensing 
in~\cite{mohammadi2026amar}}. The classifier $C_{\phi}: \mathbb{R}^{d_b} \to \mathbb{R}^{M \times (K+1)}$ outputs logits for $M$ slots with $K{+}1$ classes (where $d_b$ is the bottleneck dimension):
\begin{equation}
\hat{\mathbf{y}} = C_{\phi}(B_{\psi}(F_{\eta}(\mathbf{x}))) \in \mathbb{R}^{M \times (K+1)}.
\label{eq:multiuser_output}
\end{equation}
Each slot $m \in \{1, \ldots, M\}$ is processed with softmax to produce a 
probability distribution over $K{+}1$ classes (where class $k=K$ represents 
``no person''):
\begin{equation}
p_{m,k} = \frac{\exp(\hat{y}[m, k])}{\sum_{k'=0}^{K} \exp(\hat{y}[m, k'])}, 
\quad k \in \{0, \ldots, K\}.
\label{eq:slot_softmax}
\end{equation}
This formulation naturally represents any combination of up to $M$ 
simultaneously active users, with variable occupancy captured by slots 
predicting $\emptyset$ (class $K$) versus actual activities (classes $0, \ldots, K{-}1$).



\subsection{Hungarian Matching: Multi-User Source-Domain Training}
\label{subsec:hungarian}

During source training where ground-truth labels are available, we face a permutation invariance challenge: users can appear in any slot order, making direct index-based matching infeasible. For example, if two users perform ``walk'' and ``sit,'' the ground-truth padded set $\tilde{\mathbf{y}} = \{\text{walk}, \text{sit}, \emptyset, \ldots, \emptyset\}$ is equivalent to $\tilde{\mathbf{y}} = \{\text{sit}, \text{walk}, \emptyset, \ldots, \emptyset\}$, yet they correspond to different slot-wise labels. We employ the Hungarian algorithm to obtain optimal permutation-invariant matching {between predicted slots and ground-truth 
annotations~\cite{carion2020end, kuhn1955hungarian, mohammadi2026amar}}.
For each sample, we construct a cost matrix $\mathbf{Q} \in \mathbb{R}^{M \times M}$ where each element represents the negative log-probability of assigning predicted slot $i$ to ground-truth slot $j$:
\begin{equation}
Q_{ij} = -\log p_{i,\,\tilde{y}_j},
\label{eq:cost_matrix}
\end{equation}
where $\tilde{y}_j \in \mathcal{A} \cup \{\emptyset\}$ denotes the $j$-th element of the padded ground-truth vector $\tilde{\mathbf{y}}$, and $p_{i, \tilde{y}_j}$ is the predicted probability that slot $i$ outputs activity class $\tilde{y}_j$ (from Eq.~\ref{eq:slot_softmax}). The Hungarian algorithm efficiently finds the optimal permutation $\sigma^* \in \mathcal{S}_M$ (where $\mathcal{S}_M$ is the set of permutations over $M$ elements) that minimizes the total assignment cost:
\begin{equation}
\sigma^* = \arg\min_{\sigma \in \mathcal{S}_M} \sum_{i=1}^{M} Q_{i, \sigma(i)}.
\label{eq:hungarian_min}
\end{equation}
The source training loss uses this optimal matching to compute cross-entropy loss function as shown below:
\begin{equation}
\mathcal{L}_{\text{matched-CE}} = \frac{1}{M} \sum_{i=1}^{M} 
\mathcal{L}_{\text{CE}}(\hat{\mathbf{y}}[i], \tilde{\mathbf{y}}[\sigma^*(i)]),
\label{eq:matched_ce}
\end{equation}
where $\mathcal{L}_{\text{CE}}(\cdot, \cdot)$ is the cross-entropy loss between 
predicted logits and the ground-truth class index, and 
$\tilde{\mathbf{y}}[\sigma^*(i)] \in \{0, \ldots, K\}$ denotes the class index 
of the $\sigma^*(i)$-th element in the padded ground-truth vector. This matching-based supervision enables permutation-invariant learning during source training, i.e., the model is free to assign any activity to any slot, as the loss automatically finds the best alignment{~\cite{carion2020end, kuhn1955hungarian, mohammadi2026amar}}.


\begin{algorithm}[t]
\caption{MU-SHOT-Fi: Multi-User Adaptation}
\label{alg:multiuser}
\begin{algorithmic}[1]
\REQUIRE Pre-trained source model $f_{\theta_s} = C_{\phi_s} \circ B_{\psi_s} \circ F_{\eta_s}$, unlabeled target data $\mathcal{D}_t$, number of slots $M$, number of activity classes $K$
\ENSURE Adapted model $f_{\theta_t}$
\STATE \textbf{Stage 1: Rotation SSL Pre-training on Target Domain}
\STATE Pre-train rotation classifier $R_{\gamma}$ with frozen $F_{\eta_s}$ and $B_{\psi_s}$ (Eq.~\ref{eq:rot_loss})
\STATE \textbf{Stage 2: Joint Adaptation}
\STATE Initialize: $F_{\eta_t} \leftarrow F_{\eta_s}$, $B_{\psi_t} \leftarrow B_{\psi_s}$, $C_{\phi_t} \leftarrow C_{\phi_s}$
\STATE Freeze classifier: $C_{\phi_t}.\text{requires\_grad} = \text{False}$
\FOR{each epoch}
    \FOR{each batch $\{\mathbf{x}_i^t\}_{i=1}^{N}$ in $\mathcal{D}_t$}
        \STATE Forward: $\mathbf{y}_{\text{pred}} = C_{\phi_t}(B_{\psi_t}(F_{\eta_t}(\mathbf{x}_i^t))) \in \mathbb{R}^{N \times M \times (K+1)}$
        \STATE Compute $p_{i,m,k} = \mathrm{softmax}(\mathbf{y}_{\text{pred}}[i, m, :])$ for all $b,m$ (Eq.~\ref{eq:slot_softmax})
        \STATE \textit{// Occupancy-Weighted Information Maximization}
        \STATE Compute $\mathcal{L}_{\text{ent}}$ via Eq.~\ref{eq:entropy_multi}
        \STATE Compute $p_{\text{occ}}[i,m]$ via Eq.~\ref{eq:occupancy_prob}
        \STATE Compute $\tilde{p}_{i,m,k}$ via Eq.~\ref{eq:weighted_prob}
        \STATE Compute marginal $\bar{p}_k$ via Eq.~\ref{eq:marginal_dist}, then normalize
        \STATE Compute $\mathcal{L}_{\text{gent}}^{\text{occ}} $ via Eq.~\ref{eq:weighted_diversity}
        \STATE $\mathcal{L}_{\text{IM-multi}} = \mathcal{L}_{\text{ent}} - \mathcal{L}_{\text{gent}}^{\text{occ}}$
        \STATE \textit{// Self-Supervised Loss}
        \STATE Compute $\mathcal{L}_{\text{rot}}$ with stop-gradient (Eq.~\ref{eq:rot_loss})
        \STATE $\mathcal{L}_{\text{MU-SHOT-Fi}} = \lambda_{\text{ent}} \mathcal{L}_{\text{IM-multi}} + \lambda_{\text{rot}} \mathcal{L}_{\text{rot}}$
        \STATE Update $\{\eta_t, \psi_t, \gamma\}$ via gradient descent
    \ENDFOR
\ENDFOR
\RETURN Adapted model $f_{\theta_t}$
\end{algorithmic}
\end{algorithm}

\section{MU-SHOT-Fi: SFUDA in the Target Domain}
\label{subsec:multiuser}
MU-SHOT-Fi's SFUDA framework  combines (1) \textit{occupancy-weighted information maximization} (Section~\ref{subsubsec:occupancy_weighted_im}), which prevents collapse toward the dominant class by focusing diversity regularization on likely-occupied slots, and (2) \textit{rotation-based spatial self-supervision} (Section~\ref{subsubsec:ssl_multiuser}), which exploits CSI's frequency-time structure for domain-invariant feature learning. The classifier remains frozen while the backbone adapts.
\subsection{Occupancy-Weighted Information Maximization}
\label{subsubsec:occupancy_weighted_im}



In the source-free adaptation phase, target-domain labels are not available. As a result, the model cannot be fine-tuned using supervised cross-entropy. Instead, methods such as SHOT~\cite{liang2020we} rely on information maximization (IM), which uses prediction entropy as a self-training signal: it pushes the model toward confident decisions on target inputs while encouraging balanced use of classes through a marginal-distribution regularizer (GENT). MU-SHOT-Fi adopts this idea and designs a permutation-invariant entropy-based objective suitable for multi-user slot outputs. For a categorical distribution $\mathbf{p}$, Shannon entropy is defined as follows:

\begin{equation}
H(\mathbf{p}) = -\sum_{k} p_k \log(p_k+\epsilon),
\end{equation}
where $\epsilon$ is a small constant used for numerical stability.
High entropy indicates uncertainty (nearly uniform predictions) and low
entropy indicates confident predictions.

\subsubsection{Standard IM}
The standard IM formulation is:
\begin{equation}
\mathcal{L}_{\text{IM}} = \mathcal{L}_{\text{ent}} - \mathcal{L}_{\text{gent}}^{\text{std}},
\label{eq:im_std}
\end{equation}
where $\mathcal{L}_{\text{ent}}$ drives confident predictions and $\mathcal{L}_{\text{gent}}^{\text{std}}$ maintains class diversity via conditional and marginal entropy terms.

\paragraph{Conditional entropy minimization (confidence)}
Let $p_{i,m,k}$ be the softmax probability of class $k$ at slot $m$ for target sample $i$ in a mini-batch of size $N$, where $m\in\{1,\ldots,M\}$ and $k\in\{0,\ldots,K\}$ (with $k=K$ denoting ``no person''). Denote the slot-wise class distribution by $\mathbf{p}_{i,m}\in\mathbb{R}^{K+1}$ (Eq.~\ref{eq:slot_softmax}). We encourage confident slot-wise predictions by minimizing the average conditional entropy:
\begin{equation}
\mathcal{L}_{\text{ent}} = \frac{1}{N M}\sum_{i=1}^{N}\sum_{m=1}^{M} H(\mathbf{p}_{i,m}).
\label{eq:entropy_multi}
\end{equation}

\paragraph{GENT: Marginal entropy maximization (diversity)}
Entropy minimization alone can collapse to predicting the same class everywhere. SHOT-IM therefore uses GENT~\cite{liang2021source}, which maximizes the entropy of the batch marginal class distribution. Standard SHOT-IM computes the marginal by uniformly averaging over all $N$ samples and all $M$ slots:
\begin{equation}
\bar{p}_k^{\text{std}} = \frac{1}{N M}\sum_{i=1}^{N}\sum_{m=1}^{M} p_{i,m,k},
\quad k\in\{0,\ldots,K\},
\label{eq:standard_gent}
\end{equation}
defining the diversity loss as
\begin{equation}
\mathcal{L}_{\text{gent}}^{\text{std}} = -H(\bar{\mathbf{p}}^{\text{std}})
= \sum_{k=0}^{K}\bar{p}_k^{\text{std}}\log(\bar{p}_k^{\text{std}}+\epsilon).
\label{eq:gent_std}
\end{equation}

In multi-user sensing, Eq.~\eqref{eq:standard_gent} treats ``no person'' exactly like any activity and forces diversity over many truly-empty slots, which can amplify imbalance-driven collapse.

{To understand this failure mode formally, recall that standard information maximization (IM)~\cite{liang2020we} aims to maximize the mutual information between input $\mathbf{x}$ and prediction $\hat{\mathbf{y}}$:
\begin{equation}
I(\mathbf{x}; \hat{\mathbf{y}}) = H(\hat{\mathbf{y}}) - H(\hat{\mathbf{y}} \mid \mathbf{x}),
\end{equation}
}
{where $H(\hat{\mathbf{y}})$ promotes diversity via the marginal distribution and $H(\hat{\mathbf{y}} \mid \mathbf{x})$ enforces confident predictions. In practice, $H(\hat{\mathbf{y}})$ is approximated by the batch marginal entropy $H(\bar{\mathbf{p}}^{\text{std}})$ (Eqs.~\eqref{eq:standard_gent}--\eqref{eq:gent_std}). However, in multi-user activity classification with $M = 6$ slots and variable occupancy, the ``no person'' class $k = K$ is structurally dominant, i.e., $\bar{p}^{\text{std}}_K \gg \bar{p}^{\text{std}}_k$ for $k \in \{0, \ldots, K-1\}$. As a result, maximizing $H(\bar{\mathbf{p}}^{\text{std}})$ over all $K+1$ classes encourages uniformity including class $K$, leading to trivial collapse where predictions concentrate on the dominant class~\cite{liang2020we,liang2021source}. 
This failure mode is empirically confirmed in our experiments (Section~\ref{sec:results}), where standard IM degrades performance under domain shift.}

{The core issue is that standard IM maximizes $I(\mathbf{x}; \hat{\mathbf{y}})$ over the full label space without accounting for occupancy structure. In multi-user sensing, meaningful diversity should only be enforced over occupied slots, while empty slots should not influence the marginal distribution. This motivates our proposed formulation, which can be interpreted as maximizing the \emph{conditional} mutual information $I(\mathbf{x}; \hat{\mathbf{y}} \mid \text{occ} = 1)$, i.e., mutual information restricted to likely-occupied slots. Our occupancy-weighted GENT loss $\mathcal{L}^{\text{occ}}_{\text{gent}}$ implements this by weighting each slot's contribution with $p_{\text{occ}}[i,m] = 1 - p_{i,m,K}$: when a slot is confidently empty ($p_{i,m,K} \to 1$), its weight approaches zero and contributes nothing to diversity; when likely occupied ($p_{i,m,K} \to 0$), it contributes fully.}

{This resolves collapse, since the trivial solution of predicting $k = K$ everywhere yields zero weight for all slots and thus no diversity reward. While this shares intuition with cost-sensitive reweighting~\cite{cao2019learning}, our approach operates at the slot level within each prediction, dynamically determined by model confidence rather than dataset-level class frequencies.}

\subsubsection{Proposed Occupancy-weighted IM}
Our multi-user information maximization objective is defined as follows:
\begin{equation}
\mathcal{L}_{\text{IM-multi}} = \mathcal{L}_{\text{ent}} - \mathcal{L}_{\text{gent}}^{\text{occ}}.
\label{eq:im_multi}
\end{equation}
By focusing diversity regularization on likely-occupied slots and excluding ``no person''
from the marginal, this objective mitigates imbalance-driven collapse: occupied slots
are encouraged to spread across activities, while truly-empty slots can confidently
predict ``no person'' without being penalized for lacking diversity.
Our key modification is to keep the same \emph{confidence} term $\mathcal{L}_{\text{ent}}$
(Eq.~\eqref{eq:entropy_multi}), but redefine the \emph{diversity} term so that it
focuses on \emph{likely-occupied} slots and excludes ``no person'' from the marginal. 

First, we compute a slot occupancy probability as the probability of predicting
any activity except ``no person'', i.e.,
\begin{equation}
p_{\text{occ}}[i,m] = \sum_{k=0}^{K-1} p_{i,m,k} = 1 - p_{i,m,K}.
\label{eq:occupancy_prob}
\end{equation}
Thus, slots with high $p_{\text{occ}}[i,m]$ are likely occupied.
We then weight activity probability with the slot occupancy probability, i.e.,
\begin{equation}
\tilde{p}_{i,m,k} = p_{i,m,k}\cdot p_{\text{occ}}[i,m], \quad k=0,\ldots,K-1.
\label{eq:weighted_prob}
\end{equation}
This suppresses contributions from slots confidently predicting ``no person,'' preventing the diversity
term from pushing the model to artificially diversify among empty slots.

Next, we form the \emph{occupancy-weighted marginal} over activity classes only (excluding ``no person''), i.e.,
\begin{equation}
\bar{p}_k^{\text{occ}} =
\frac{1}{Z}\cdot\frac{1}{N M}\sum_{i=1}^{N}\sum_{m=1}^{M}\tilde{p}_{i,m,k},
\quad k=0,\ldots,K-1,
\label{eq:marginal_dist}
\end{equation}
where $Z=\sum_{k=0}^{K-1}\bar{p}_k^{\text{occ}}$ ensures normalization.
We then define the corresponding GENT loss function term as follows:
\begin{equation}
\mathcal{L}_{\text{gent}}^{\text{occ}} = -H(\bar{\mathbf{p}}^{\text{occ}})
= \sum_{k=0}^{K-1}\bar{p}_k^{\text{occ}}\log(\bar{p}_k^{\text{occ}}+\epsilon).
\label{eq:weighted_diversity}
\end{equation}

\subsection{Rotation-based Spatial Self-Supervision}
\label{subsubsec:ssl_multiuser}

To align source and target representations without labels, we add self-supervised
learning (SSL) losses on the shared CSI input and a shared bottleneck embedding,
rather than on individual slot outputs. Concretely, let
\[
\mathbf{b}(\mathbf{x}) = B_{\psi}(F_{\eta}(\mathbf{x})) \in \mathbb{R}^{d_b},
\]
denotes the bottleneck feature for sample $\mathbf{x}$. SSL regularizes this shared
embedding, benefiting all slots simultaneously.

Specifically, we use a rotation prediction task ($0^\circ$ vs.\ $180^\circ$) on the
frequency--time CSI grid~\cite{gidaris2018unsupervised}. Let
$\mathrm{rot}(\mathbf{x}, r)$ be $\mathbf{x}$ rotated by angle $r \in \{0^\circ,180^\circ\}$.
The rotation head $R_{\gamma}$ takes the concatenation of bottleneck features from
the original and rotated inputs and predicts $r$, i.e.,
\[
R_{\gamma}\!\left([\mathbf{b}(\mathbf{x});\mathbf{b}(\mathrm{rot}(\mathbf{x},r))]\right)\in\mathbb{R}^2.
\]

Following SHOT++~\cite{liang2021source}, we (i) pre-train $R_{\gamma}$ on target
data while freezing the source feature extractor, then (ii) keep the task during
adaptation but apply stop-gradient on the original branch to avoid interfering
with activity classification:
\begin{equation}
\mathcal{L}_{\text{rot}}
=
\mathbb{E}_{\mathbf{x}, r}
\Big[
\mathcal{L}_{\text{CE}}\big(
R_{\gamma}(
[\mathrm{sg}(\mathbf{b}(\mathbf{x}));
 \mathbf{b}(\mathrm{rot}(\mathbf{x},r))]
),
r
\big)
\Big].
\label{eq:rot_loss}
\end{equation}
Here $\mathrm{sg}(\cdot)$ denotes stop-gradient~\cite{grill2020bootstrap}, and
$[\cdot;\cdot]$ denotes concatenation. Stop-gradient on the original branch
prevents the rotation loss from pulling the shared embedding away from what the
main slot-prediction objective needs, while the rotated branch still provides
spatial regularization.

\subsection{Proposed Loss Function}
The proposed multi-user adaptation objective combines occupancy-weighted information maximization and rotation-based self-supervised loss function and can be formulated as:
\begin{equation}
\mathcal{L}_{\text{MU-SHOT-Fi}}
=
\lambda_{\text{ent}}\mathcal{L}_{\text{IM-multi}}
+
\lambda_{\text{rot}}\mathcal{L}_{\text{rot}}.
\label{eq:total_loss_multiuser}
\end{equation}
where $\lambda_{\text{ent}}$ and $\lambda_{\text{rot}}$ represent the weight of contribution of each loss term. Training proceeds in two stages: (i) rotation SSL pre-training on unlabeled target data with frozen source features, followed by (ii) joint adaptation with
$\mathcal{L}_{\text{IM-multi}}$ and $\mathcal{L}_{\text{rot}}$. Algorithm~\ref{alg:multiuser} details the sequence of training procedure.

\begin{figure*}
    \centering
    \includegraphics[width=0.90\linewidth]{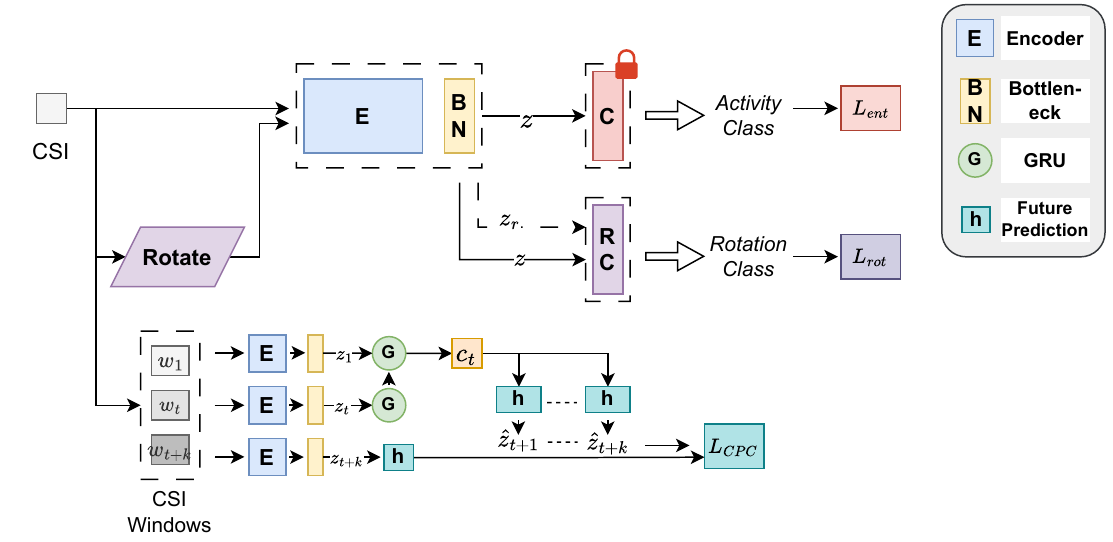}
    \caption{Architecture of the proposed SU-SHOT-Fi source-free unsupervised domain adaptation framework for single-user Wi-Fi HAR. The backbone is updated through (i) rotation-based spatial self-supervision exploiting CSI frequency-time structure and (ii) contrastive predictive coding (CPC) capturing temporal consistency across CSI windows, while the classifier remains frozen during target adaptation.}
    \label{fig:stshot_singleuser}
\end{figure*}

\begin{algorithm}[t]
\caption{SU-SHOT-Fi: Single-User Adaptation}
\label{alg:single_user}
\begin{algorithmic}[1]
\REQUIRE Pre-trained source model $f_{\theta_s} = C_{\phi_s} \circ B_{\psi_s} \circ F_{\eta_s}$, unlabeled target data $\mathcal{D}_t$, number of activity classes $K$
\ENSURE Adapted model $f_{\theta_t}$
\STATE \textbf{Stage 1: SSL Pre-training on Target Domain}
\STATE Pre-train rotation classifier $R_{\gamma}$ with frozen $F_{\eta_s}$, $B_{\psi_s}$ (Eq.~\ref{eq:rot_loss})
\STATE Pre-train CPC model $(g_{\xi}, h_{\omega}, \{W_k\})$ on $\mathcal{D}_t$ (Eq.~\ref{eq:cpc_loss})
\STATE \textbf{Stage 2: Joint Adaptation with Pseudo-Labeling}
\STATE Initialize: $F_{\eta_t} \leftarrow F_{\eta_s}$, $B_{\psi_t} \leftarrow B_{\psi_s}$, $C_{\phi_t} \leftarrow C_{\phi_s}$; Freeze $C_{\phi_t}$
\FOR{each epoch}
    \STATE \textit{// Generate Pseudo-Labels via K-Nearest Centroids}
    \STATE Extract $\mathbf{z}_j = B_{\psi_t}(F_{\eta_t}(\mathbf{x}_j^t))$, $\mathbf{p}_j = \sigma(C_{\phi_t}(\mathbf{z}_j))$ for all $j \in \{1, \ldots, n_t\}$
    \STATE Normalize: $\tilde{\mathbf{z}}_j = [\mathbf{z}_j; 1] / \|[\mathbf{z}_j;1]\|_2$
    \STATE Initialize centroids: $\mathbf{c}_k = \sum_{j=1}^{n_t} [\mathbf{p}_j]_k \tilde{\mathbf{z}}_j \big/ \sum_{j=1}^{n_t} [\mathbf{p}_j]_k$, $k \in \{1, \ldots, K\}$
    \STATE Assign pseudo-labels: $\hat{y}_j = \arg\min_{k} d_{\text{cosine}}(\tilde{\mathbf{z}}_j, \mathbf{c}_k)$; refine centroids once
    \FOR{each batch $\{\mathbf{x}_j^t\}$}
        \STATE Compute $\mathcal{L}_{\text{PL}}$ (Eq.~\ref{eq:pseudo_label}), $\mathcal{L}_{\text{IM}}$ (Eq.~\ref{eq:im_std}), $\mathcal{L}_{\text{rot}}$ (Eq.~\ref{eq:rot_loss}), $\mathcal{L}_{\text{CPC}}$ (Eq.~\ref{eq:cpc_loss})
        \STATE $\mathcal{L}_{\text{SU-SHOT-Fi}} = \lambda_{\text{cls}} \mathcal{L}_{\text{PL}} + \lambda_{\text{ent}} \mathcal{L}_{\text{IM}} + \lambda_{\text{rot}} \mathcal{L}_{\text{rot}} + \lambda_{\text{cpc}} \mathcal{L}_{\text{CPC}}$
        \STATE Update $\{\eta_t, \psi_t, \gamma, \xi, \omega, \{W_k\}\}$ via gradient descent
    \ENDFOR
\ENDFOR
\RETURN Adapted model $f_{\theta_t}$
\end{algorithmic}
\end{algorithm}

\section{SU-SHOT-Fi: Single-User HAR as a Special Case}
\label{subsec:singleuser}

The MU-SHOT-Fi framework simplifies naturally to single-user scenarios where each CSI sample corresponds to exactly one activity from $K$ mutually exclusive classes (see Algorithm~\ref{alg:single_user} and 
Fig.~\ref{fig:stshot_singleuser}). In this setting, the proposed adaptation strategy  for SU-SHOT-Fi is explained below.

\subsection{Standard IM with no Hungarian Matching} First, occupancy is fixed at one user per sample, eliminating variable occupancy. Thus, we use the standard SHOT-IM objective~\cite{liang2020we} denoted by $\mathcal{L}_{\text{IM}}$, which combines
(i) \emph{conditional entropy minimization} to encourage confident predictions, and (ii) \emph{marginal entropy maximization} (GENT) to avoid degenerate collapse by promoting diverse classes across the target batch. Second, with only one prediction per sample, slot permutation invariance disappears-there is no need for Hungarian matching as predictions map directly to ground truth during source training. 

\subsection{Clustering-based Pseudo-labeling} Unlike multi-user scenarios, single-user datasets exhibit relatively balanced class distributions across activities~\cite{zhang2021widar3}. This balanced setting enables SHOT's clustering-based pseudo-labeling~\cite{liang2020we}, since the model won't converge to the dominant class which is a limitation of multi-user case.
Pseudo-labeling generates supervision through clustering. At each epoch, we extract bottleneck features $\mathbf{z}_j = B_{\psi}(F_{\eta}(\mathbf{x}_j^t)) \in \mathbb{R}^{d_b}$ and predictions $\mathbf{p}_j = \sigma(C_{\phi}(\mathbf{z}_j)) \in \mathbb{R}^K$ for all target samples $j \in \{1, \ldots, n_t\}$. Features are normalized with a bias term 
$\tilde{\mathbf{z}}_j = [\mathbf{z}_j; 1] / \|\mathbf{z}_j\|_2 \in \mathbb{R}^{d_b+1}$, and class centroids are initialized as the prediction-weighted average of normalized features, i.e.,
\begin{equation}
\mathbf{c}_k = \frac{\sum_{j=1}^{n_t} p_{jk} \tilde{\mathbf{z}}_j}{\sum_{j=1}^{n_t} p_{jk}}, 
\quad k \in \{1, \ldots, K\}.
\label{eq:centroid_init}
\end{equation}
Each sample is then assigned to its nearest centroid using cosine distance:
\begin{equation}
\hat{y}_j = \arg\min_{k \in \{1,\ldots,K\}} d_{\text{cosine}}(\tilde{\mathbf{z}}_j, \mathbf{c}_k),
\label{eq:pseudo_assignment}
\end{equation}
where $d_{\text{cosine}}(\mathbf{u}, \mathbf{v}) = 1 - \frac{\mathbf{u}^\top \mathbf{v}}{\|\mathbf{u}\|_2 \|\mathbf{v}\|_2}$. Centroids are refined once using the following assignments, i.e.,
\begin{equation}
\mathbf{c}_k \leftarrow \frac{\sum_{j: \hat{y}_j = k} \tilde{\mathbf{z}}_j}{|\{j: \hat{y}_j = k\}|},
\label{eq:centroid_refine}
\end{equation}
and pseudo-labels are then reassigned. The pseudo-labeling loss supervises the model, i.e.,
\begin{equation}
\mathcal{L}_{\text{PL}} = \frac{1}{n_t} \sum_{j=1}^{n_t} \mathcal{L}_{\text{CE}}(C_{\phi}(B_{\psi}(F_{\eta}(\mathbf{x}_j^t))), \hat{y}_j).
\label{eq:pseudo_label}
\end{equation}
The classifier outputs $C_{\phi}: \mathbb{R}^{d_b} \to \mathbb{R}^{K}$ logits over $K$ activities without the ``no person'' class. 

\subsection{Temporal SSL via CPC}
We also consider Contrastive Predictive Coding (CPC)~\cite{oord2018representation} to exploit temporal structure in CSI sequences. While rotation-based spatial SSL targets invariances in the frequency--time grid, temporal dynamics (e.g., periodic stride patterns during walking) provide domain-invariant cues even when absolute CSI values shift across environments and hardware.
CPC splits $\mathbf{x}\in\mathbb{R}^{1\times F\times T}$ into $W=\lfloor T/w\rfloor$ windows of length $w$, encodes each window with encoder $g_{\xi}$, summarizes past context with GRU $h_{\omega}$, and predicts future embeddings using InfoNCE loss:
\begin{equation}
\mathcal{L}_{\text{CPC}}
=
\frac{1}{K_p}\sum_{k=1}^{K_p}
\left[
-\frac{1}{N}\sum_{i=1}^{N}
\log
\frac{\exp(s_{i,i}^{(k)})}{\sum_{j=1}^{N}\exp(s_{i,j}^{(k)})}
\right],
\label{eq:cpc_loss}
\end{equation}
where $K_p$ is the prediction horizon and $s_{i,j}^{(k)}=\hat{\mathbf{z}}_{t+k}^{(i)\top}\mathbf{z}_{t+k}^{(j)}/\tau$ is the temperature-scaled cosine similarity between predicted future embedding $\hat{\mathbf{z}}_{t+k}^{(i)}$ and true future embedding $\mathbf{z}_{t+k}^{(j)}$.

\subsection{SU-SHOT-Fi Loss Function:} The proposed SU-SHOT-Fi objective can then be given as:
\begin{equation}
\mathcal{L}_{\text{SU-SHOT-Fi}} = \lambda_{\text{cls}} \mathcal{L}_{\text{PL}} 
+ \lambda_{\text{ent}} \mathcal{L}_{\text{IM}} + \lambda_{\text{rot}} 
\mathcal{L}_{\text{rot}} + \lambda_{\text{cpc}} \mathcal{L}_{\text{CPC}},
\label{eq:total_loss}
\end{equation}
where $\mathcal{L}_{\text{PL}}$ is the pseudo-labeling loss, $\mathcal{L}_{\text{IM}}$ is standard information maximization, and $\lambda_{\text{cls}}$, $\lambda_{\text{cpc}}$ are the corresponding loss weights. Training follows the same two-stage procedure: SSL pre-training of the rotation classifier and CPC model on unlabeled target data, followed by joint adaptation combining SHOT-IM with CSI-specific self-supervised losses. Algorithm~\ref{alg:single_user} summarizes the complete procedure, and Fig.~\ref{fig:stshot_singleuser} illustrates the architecture. As shown in Section~\ref{subsec:cpc_analysis}, CPC benefits single-user scenarios but conflicts with multi-user slot-level predictions due to permutation invariance requirements.

\section{Experimental Set-Up, Baselines, and Evaluation Metrics}
\label{sec:experiments}


\subsection{Considered Datasets}
\label{subsec:datasets}
We validate MU-SHOT-Fi on two datasets: WiMANS~\cite{huang2024wimans} for multi-user activity recognition under environmental and hardware variations, and Widar 3.0~\cite{zhang2021widar3} for single-user gesture recognition under environmental and positional shifts. 
\subsubsection{WiMANS~\cite{huang2024wimans}} is a comprehensive multi-user dataset captured in classroom and meeting room environments at both 2.4 GHz and 5 GHz frequencies using a 3×3 MIMO configuration. The dataset contains samples with 0 to 5 simultaneous users performing 9 distinct activities: nothing, walk, rotation, jump, wave, lie down, pick up, sit down, and stand up. Each CSI measurement is collected at 1000 Hz sampling rate across 30 subcarriers over a 3-second window, yielding complex-valued tensors $\mathbf{x}_{\text{raw}} \in \mathbb{C}^{3000 \times 30 \times 3 \times 3}$ representing temporal samples, subcarriers, transmit antennas, and receive antennas, respectively. Following standard pre-processing~\cite{huang2024wimans}, we extract amplitude and reshape to $\mathbf{x} \in \mathbb{R}^{1 \times 3000 \times 270}$ where the spatial dimensions are flattened (30 subcarriers × 3 transmit × 3 receive = 270 features). To test MU-SHOT-Fi, we take $M=6$ and $K+1=10$ classes (9 activities plus "no person"). 

The source domain consists of classroom samples at 2.4 GHz. We evaluate three domain shift scenarios: 

(1) \textbf{Cross-Room} (classroom to meeting room at 2.4 GHz) evaluates performance under  environmental changes, such as room geometry and furniture layout changes; 

(2) \textbf{Cross-Frequency} (classroom at 2.4 GHz to 5 GHz) evaluates performance under frequency changes. 

(3) \textbf{Combined Shift} (classroom at 2.4 GHz to meeting room at 5 GHz) evaluates performance under simultaneous environmental and frequency changes.

\subsubsection{Widar 3.0~\cite{zhang2021widar3}} contains gesture samples 
across 6 classes (push and pull, sweep, clap, slide, draw-O(H), draw-zigzag(H)) from 9 users across 5 physical locations and 5 body orientations, collected at 5.825~GHz with one transmitter and six receivers (3 antennas each). Raw CSI has dimensions $(N_t \times N_r \times N_s \times T) = (1 \times 18 \times 30 \times \text{variable})$ 
for transmit antenna, receive antennas (6 receivers × 3 antennas), subcarriers, and temporal samples.
Following standard preprocessing~\cite{yan2025wi}, we standardize temporal length to 1200 samples and extract CSI phase ratios. Raw CSI $\mathbf{x}_{\text{raw}} \in \mathbb{C}^{1 \times 18 \times 30 \times T}$ 
represents 1 transmit antenna, 18 receive antennas (6 receivers × 3 antennas per receiver), 30 subcarriers, and temporal samples. For each receiver $r \in \{1, \ldots, 6\}$, we compute the CSI ratio between its first two antennas $H_r^{\text{ratio}}(k,t) = H_{r,1}(k,t) / H_{r,2}(k,t)$ for subcarrier $k$ and time $t$, which eliminates time-varying phase offsets~\cite{yan2025wi}. We extract the phase $\phi_r(k,t) = \angle H_r^{\text{ratio}}(k,t)$ and stack across receivers and subcarriers, yielding $\mathbf{x} \in \mathbb{R}^{1 \times 180 \times 1200}$ where $180 = 6 \text{ receivers} \times 30 \text{ subcarriers}$. We evaluate three domain shift scenarios: 

(1) \textbf{Cross-Room} (Room 1 to Room 2) evaluates environmental shift from room geometry and materials; 

(2) \textbf{Cross-Torso} (orientations 2–5 to 1) evaluates body rotation within the same room; 

(3) \textbf{Cross-Face} (locations 2–5 to 1) evaluates head orientation changes affecting signal shadowing.

{WiMANS serves as the primary benchmark for 
evaluating MU-SHOT-Fi, as it provides multi-user CSI samples with 
simultaneous activity annotations across diverse domain shift 
conditions (cross-room, cross-frequency, and combined). To the best 
of our knowledge, it is the only open-access dataset offering this 
combination~\cite{radwan2025tutorial, wang2026survey}, making it 
the natural testbed for all multi-user claims, including 
occupancy-weighted information maximization, permutation-invariant 
set prediction, and dominant-class collapse prevention. Widar~3.0 
complements this evaluation by validating the core adaptation 
components shared between MU-SHOT-Fi and SU-SHOT-Fi, namely 
rotation-based self-supervision and information maximization, in a 
controlled single-user setting where confounding factors such as 
signal entanglement and variable occupancy are absent.}

\begin{table}[t]
\centering
\caption{WiMANS hyperparameters for MU-SHOT-Fi}
\label{tab:wimans_hyperparams}
\small
\resizebox{\linewidth}{!}{%
\begin{tabular}{@{}llc@{}}
\toprule
\textbf{Component} & \textbf{Parameter} & \textbf{Value} \\
\midrule
\multirow{4}{*}{General} 
& Adaptation epochs & 50 \\
& Batch size & 64 \\
& Optimizer & Adam \\
& Adaptation learning rate & $1 \times 10^{-4}$ \\
\midrule
\multirow{6}{*}{\shortstack[l]{Multi-User\\Backbone}}
& Slots ($M$) & 6 \\
& Classes per slot & 10 \\
& Label smoothing ($\epsilon$) & 0.2 \\
& Entropy minimization ($\lambda_{\text{ent}}$) & 1.0 \\
& Diversity maximization & Enabled \\
& Pseudo-labeling weight ($\lambda_{\text{cls}}$) & 0.0 \\
\midrule
\multirow{2}{*}{Rotation SSL}
& Pre-training epochs & 70 \\
& Loss weight ($\lambda_{\text{rot}}$) & 0.5 \\
\bottomrule
\end{tabular}
}
\end{table}

\begin{table}[t]
\centering
\caption{Widar 3.0 hyperparameters for SHOT-Fi}
\label{tab:widar_hyperparams}
\small
\resizebox{\linewidth}{!}{%
\begin{tabular}{@{}llc@{}}
\toprule
\textbf{Component} & \textbf{Parameter} & \textbf{Value} \\
\midrule
\multirow{4}{*}{General} 
& Adaptation epochs & 70 \\
& Batch size & 32 \\
& Optimizer & Adam \\
& Adaptation learning rate & $1 \times 10^{-4}$ \\
\midrule
\multirow{3}{*}{SHOT-IM Backbone}
& Pseudo-labeling weight ($\lambda_{\text{cls}}$) & 0.1 \\
& Entropy minimization ($\lambda_{\text{ent}}$) & 1.0 \\
& Diversity maximization & Enabled \\
\midrule
\multirow{2}{*}{Rotation SSL}
& Pre-training epochs & 70 \\
& Loss weight ($\lambda_{\text{rot}}$) & 0.3 \\
\midrule
\multirow{11}{*}{CPC}
& Pre-training epochs & 70 \\
& Pre-training learning rate & $1 \times 10^{-3}$ \\
& Loss weight ($\lambda_{\text{cpc}}$) & 0.3 \\
& Window size ($w$) & 10 timesteps \\
& Prediction steps ($K_p$) & 9 \\
& Encoder embedding dim ($d_e$) & 256 \\
& Context (GRU) hidden dim ($d_c$) & 512 \\
& Projection head dim ($d_p$) & 256 \\
& InfoNCE temperature ($\tau$) & 0.07 \\
& Mask probability & 0.5 \\
& Mask ratio (per window) & 0.15 \\
\bottomrule
\end{tabular}
}
\end{table}
\subsection{Evaluation Metrics}
\label{subsec:metrics}
We report slot-level results after Hungarian alignment (slot-wise accuracy,  Activity Macro-F1), sample-level performance (exact match accuracy), and occupancy-level performance (occupancy MAE, occupancy exact match).

\subsubsection{Multi-User HAR Evaluation Metrics}
Each test sample contains $M=6$ user slots with $M$ categorical predictions and $M$ ground-truth slot labels, where a dedicated \emph{no-person} class denotes an empty slot. Since user ordering is arbitrary, we align predicted slots with ground-truth slots using Hungarian matching~\cite{kuhn1955hungarian} given by:
\begin{equation}
\pi^\star
=
\arg\min_{\pi \in \mathcal{S}_M}
\sum_{m=1}^{M}
-\log p\!\left(\hat{y}_{i,m}=y_{i,\pi(m)}\right),
\end{equation}
where $\mathcal{S}_M$ is the set of permutations over $M$ slots and $p(\cdot)$ is the model's softmax probability. 

\paragraph{Slot-wise Accuracy} We then compute slot-wise accuracy as follows:
\begin{equation}
\text{SlotAcc}
=
\frac{1}{N M}
\sum_{i=1}^{N}
\sum_{m=1}^{M}
\mathbb{I}\!\left(\hat{y}_{i,m}=y_{i,\pi^\star(m)}\right).
\end{equation}

\paragraph{Activity Macro-F1} evaluates recognition quality over the $K$ activity classes only (excluding $\emptyset$). Let $\mathcal{I}=\{(i,m): y_{i,\pi^\star(m)}\neq \emptyset\}$ denotes occupied ground-truth slots:
\begin{equation}
\text{ActivityF1}
=
\frac{1}{K}
\sum_{k=1}^{K}
\text{F1}\Big(\{\hat{y}_{i,m}\}_{(i,m)\in\mathcal{I}}, \{y_{i,\pi^\star(m)}\}_{(i,m)\in\mathcal{I}}; k\Big).\nonumber
\end{equation}

\paragraph{Exact Match Accuracy}
Exact Match counts a prediction as correct only if all $M$ slots match after Hungarian alignment~{\cite{mohammadi2026amar}}, i.e.,
\begin{equation}
\text{ExactMatch}
=
\frac{1}{N}
\sum_{i=1}^{N}
\mathbb{I}\!\left(
\bigwedge_{m=1}^{M}
\hat{y}_{i,m}=y_{i,\pi^\star(m)}
\right).
\end{equation}

\paragraph{Occupancy Metrics}
Ground-truth and predicted occupancy for sample $i$ can be defined as $o_i = \sum_{m=1}^{M} \mathbb{I}(y_{i,m}\neq \emptyset)$ and $\hat{o}_i = \sum_{m=1}^{M} \mathbb{I}(\hat{y}_{i,m}\neq \emptyset)$, respectively. The occupancy metrics are then given as follows:
\begin{equation}
\text{OccMAE}
=
\frac{1}{N}
\sum_{i=1}^{N}
\left|o_i-\hat{o}_i\right|,
\quad
\text{OccExact}
=
\frac{1}{N}
\sum_{i=1}^{N}
\mathbb{I}(o_i=\hat{o}_i).\nonumber
\end{equation}

\textit{Remark:} ExactMatch is the strictest end-to-end metric, requiring all $M$ aligned slots to match. ActivityF1 focuses on activity recognition over occupied slots, reducing the influence of the no-person class $\varnothing$, while SlotAcc (computed over all matched slot labels including $\varnothing$) can become inflated when many slots are empty.

{\textit{Remark on metric interdependence:} Activity F1 is computed exclusively over ground-truth occupied slots and does not penalize predictions on empty slots. Consequently, a model that over-predicts activities across all $M$ slots can achieve relatively high Activity F1 while incurring large occupancy errors, as the metric rewards any correct activity match regardless of occupancy accuracy.}

{Conversely, improvements in Slot-wise Accuracy and Occupancy MAE that arise from more precise occupancy estimation may coincide with lower Activity F1, since fewer but more selective activity predictions reduce the chance of incidental matches in occupied slots.}

{Therefore, no single metric is sufficient for evaluating multi-user performance: Activity F1 must be interpreted jointly with occupancy metrics to distinguish genuine activity recognition from over-prediction, and Slot-wise Accuracy should be considered alongside Activity F1 to verify that gains reflect activity discrimination rather than empty-slot exploitation.}

\subsubsection{Single-User HAR Evaluation Metrics}
For single-user gesture recognition, the evaluation simplifies to standard multi-class classification.
Unlike multi-user scenarios where variable occupancy and class imbalance create fundamental evaluation challenges, single-user metrics follow established classification conventions. 
Each sample corresponds to exactly one gesture from $K=6$ classes, and we report:
\paragraph{Classification Accuracy} The fraction of correctly classified samples, i.e.,
$
\text{Accuracy} = \frac{1}{N} \sum_{i=1}^{N} \mathbb{I}(\hat{y}_i = y_i),
$
where $\hat{y}_i$ is the predicted class and $y_i$ is the ground-truth class.

\paragraph{Per-Class F1-Score} To evaluate performance across gesture types beyond overall accuracy, we compute F1-scores for each class $k$ using standard binary classification metrics, i.e.,
$
\text{F1}_k = \frac{2 \cdot \text{Precision}_k \cdot \text{Recall}_k}{\text{Precision}_k + \text{Recall}_k}.
$
We report both per-class F1-scores and their macro-average. Unlike accuracy, macro-F1 weights each gesture equally and highlights gesture-specific precision/recall failures, revealing which gestures are most affected by domain shift and whether adaptation improves performance uniformly across classes.

\subsection{Considered Baselines}
\label{subsec:baselines}

We compare MU-SHOT-Fi and SU-SHOT-Fi against: 

(1) \textbf{Source-only} model trained on labeled source data and tested on target data with no adaptation, to quantify the domain-shift gap; 

(2) \textbf{SHOT-IM~\cite{liang2020we}}, a standard source-free UDA baseline based on information maximization, with pseudo-labeling enabled only in the single-user case; 

(3) \textbf{SHOT++~\cite{liang2021source}}, an improved SHOT-IM variant with additional SSL target-side regularization, with pseudo-labeling enabled only in the single-user case.

We select these baselines because, to the best of our knowledge, there are no prior SFUDA methods tailored to multi-user Wi-Fi sensing; thus, \textit{Source-only} quantifies the domain-shift gap and SHOT-IM/SHOT++~\cite{liang2020we,liang2021source} serves as the closest and strongest existing SFUDA references for assessing how standard single-user adaptation objectives behave when extended to permutation-invariant, variable-occupancy outputs.

For a fair comparison in the multi-user scenario, all methods use the same set-prediction formulation and Hungarian matching during source training to handle permutation invariance. During target adaptation, pseudo-labeling is disabled for SHOT-IM/SHOT++ in the multi-user setting since slot-level pseudo-labels are ill-defined without correspondence and become unstable under variable occupancy and dominant $\emptyset$ slots.
For SHOT++ we also disable the MixMatch stage since it is a generic semi-supervised add-on that could be applied to all methods~\cite{{berthelot2019mixmatch}}. 
We also evaluate an architectural variant for the multi-user set predictor by replacing the parallel classification heads with a DETR-style transformer decoder with learned query embeddings~\cite{carion2020end}. This comparison isolates the effect of the set-prediction backbone, while keeping the adaptation procedure identical (Hungarian matching, occupancy-weighted GENT, and rotation SSL) for a fair assessment.

\begin{table*}[t]
\centering
\caption{WiMANS: SFUDA results under three domain shifts. Values are mean$\pm$std across runs. Higher is better ($\uparrow$) except where noted ($\downarrow$). Best within each scenario is \textbf{bolded} and second-best is \underline{underlined}.}
\label{tab:wimans_vertical}
\scriptsize
\setlength{\tabcolsep}{3pt}
\renewcommand{\arraystretch}{1.15}
\begin{tabular}{llccccc}
\toprule
Scenario & Method &
Exact Match (\%) $\uparrow$ &
Activity F1 (\%)\textsuperscript{\dag} $\uparrow$ &
Slot-wise Acc (\%) $\uparrow$ &
Occupancy MAE $\downarrow$ &
Occupancy Exact Match (\%) $\uparrow$ \\
\midrule
\multirow{5}{*}{Cross-Room}
& Source Only & 2.12$\pm$0.43 & 15.89$\pm$0.47 & {55.09$\pm$2.47} & \textbf{1.52$\pm$0.24} & 20.45$\pm$4.77 \\
& SHOT-IM       & 2.82$\pm$2.94 & \textbf{24.97$\pm$1.01} & 37.12$\pm$7.09 & 2.07$\pm$0.47 &  {17.99$\pm$7.68} \\
& SHOT++     & \underline{7.05$\pm$2.17} & {16.88$\pm$1.41} & \underline{55.32$\pm$1.96} & {1.63$\pm$0.26} & \underline{23.28$\pm$4.86}  \\
& MU-SHOT-Fi (Ours)       & \textbf{7.76$\pm$1.51} & \underline{17.08$\pm$1.15} & {55.29$\pm$1.98} & \underline{1.60$\pm$0.25} & \textbf{24.61$\pm$3.72}  \\
& MU-SHOT-Fi + CPC & 6.34$\pm$2.41 & 16.16$\pm$0.54 & \textbf{55.43$\pm$1.37} & 1.63$\pm$0.25 & 22.75$\pm$4.98 \\
\midrule

\multirow{5}{*}{Cross-Frequency}
& Source Only & 0.00$\pm$0.00 & \underline{25.69$\pm$0.97} & 21.16$\pm$4.71 & 2.92$\pm$0.33 & 7.40$\pm$1.14  \\
& SHOT-IM       & 0.35$\pm$0.49 & \textbf{26.56$\pm$2.86} & 32.39$\pm$2.91 & 2.61$\pm$0.32 & 9.34$\pm$2.17  \\
& SHOT++     & \underline{0.71$\pm$0.25} & 19.61$\pm$1.59 & \textbf{47.08$\pm$1.57} & \underline{1.89$\pm$0.16} & \textbf{15.16$\pm$3.52}  \\
& MU-SHOT-Fi (Ours)   & \underline{0.71$\pm$0.25} & 18.82$\pm$1.50 & \underline{46.97$\pm$1.35} & \textbf{1.89$\pm$0.15} & \underline{14.63$\pm$3.52} \\
& MU-SHOT-Fi + CPC & \textbf{1.41$\pm$0.24} & 19.55$\pm$2.13 & \textbf{47.33$\pm$1.91} & 1.91$\pm$0.14 & 13.40$\pm$2.87 \\
\midrule

\multirow{5}{*}{Combined Shift}
& Source Only & 0.00$\pm$0.00 & \textbf{24.99$\pm$2.53} & 19.61$\pm$4.08 & 3.00$\pm$0.35 & 5.99$\pm$4.75  \\
& SHOT-IM       & {0.00$\pm$0.00} & \underline{23.62$\pm$2.75} & 28.63$\pm$4.21 & 2.80$\pm$0.38 & 9.17$\pm$4.11 \\
& SHOT++     & \textbf{0.18$\pm$0.25} & 19.23$\pm$1.94 & {41.06$\pm$2.76} & {2.16$\pm$0.16} & \underline{10.58$\pm$2.62} \\
& MU-SHOT-Fi (Ours)     & \textbf{0.18$\pm$0.25} & 18.47$\pm$2.91 & \textbf{41.97$\pm$2.39} & \textbf{2.12$\pm$0.12} &  \textbf{10.93$\pm$1.79}  \\
& MU-SHOT-Fi + CPC & \textbf{0.18$\pm$0.25} & 18.45$\pm$2.41 & \underline{41.58$\pm$2.68} & \underline{2.15$\pm$0.14} & 5.99$\pm$4.75 \\
\midrule

\multirow{5}{*}{\textbf{Average}}
& Source Only & 0.71$\pm$0.14 & \underline{22.19$\pm$1.32} & 31.95$\pm$3.75 & 2.48$\pm$0.31 & 11.28$\pm$3.55 \\
& SHOT-IM       & 1.06$\pm$1.14 & \textbf{25.05$\pm$2.21} & 32.71$\pm$4.74 & 2.49$\pm$0.39 & 12.17$\pm$4.65 \\
& SHOT++     & \underline{2.65$\pm$0.89} & 18.57$\pm$1.65 & {47.82$\pm$2.10} & \underline{1.89$\pm$0.19} & \underline{16.34$\pm$3.67} \\
& MU-SHOT-Fi (Ours) & \textbf{2.88$\pm$0.67} & {18.12$\pm$1.85} & \underline{48.08$\pm$1.91} & \textbf{1.87$\pm$0.17} & \textbf{16.72$\pm$3.01} \\

& MU-SHOT-Fi + CPC & 2.64$\pm$0.97 & 18.05$\pm$1.69 & \textbf{48.11$\pm$1.99} & 1.90$\pm$0.18 & 14.04$\pm$4.20 \\
\midrule
\multicolumn{2}{l}{{Random Predictor}} & {$\approx$0.00} & {$\approx$11.11} & {$\approx$10.00} & {$\approx$5.00} & {$\approx$1.54} \\
\bottomrule
\vspace{-1mm}
\end{tabular}
\small

Note: Average row shows mean performance across the three domain shift scenarios. {Random Predictor assumes uniform predictions over $K+1=10$ classes per slot.}
\end{table*}

\subsection{Model Hyperparameter Settings}
\subsubsection{MU-SHOT-Fi} The feature extractor consists of three 2D convolutional blocks with kernels (27,27), (15,15), (7,7); strides (7,7), (3,3), (1,1); and channels 32, 64, 128. Each block uses batch normalization, LeakyReLU activation, and dropout (0.2). Global average pooling produces 128-dimensional features compressed through a bottleneck to 128 dimensions. The classifier outputs $6 \times 10$ logits for $M=6$ slots with $K+1=10$ classes. Source model trains for 50 epochs with Hungarian matching, label smoothing ($\epsilon=0.2$), and learning rate $1 \times 10^{-3}$. Adaptation runs for 50 epochs with batch size 64, learning rate $1 \times 10^{-4}$, and loss weights $\lambda_{\text{ent}}=1.0$ and $\lambda_{\text{rot}}=0.5$. Complete hyperparameters are given in Table~\ref{tab:wimans_hyperparams}.

\subsubsection{SU-SHOT-Fi} We use ResNet-18 as feature extractor, modified for input shape $(B, 1, 180, 1200)$. The bottleneck reduces features to dimension 512 and the classifier outputs logits for 6 classes. Source models train for 30 epochs with label smoothing ($\epsilon=0.1$), SGD (learning rate 0.1, momentum 0.9, weight decay $5 \times 10^{-4}$), and batch size 32. CPC uses window size $w=10$ (120 windows per 1200-timestep sample), predicts $K=9$ future windows, with encoder dimension 256, GRU hidden dimension 512, and InfoNCE temperature $\tau=0.07$. Adaptation runs for 70 epochs with Adam optimizer, learning rate $1 \times 10^{-4}$, batch size 32, and loss weights $\lambda_{\text{cls}}=0.1$, $\lambda_{\text{ent}}=1.0$, $\lambda_{\text{rot}}=0.3$, $\lambda_{\text{cpc}}=0.3$. Results average over 3 runs with different random seeds. Complete hyperparameters are given in Table~\ref{tab:widar_hyperparams}.

\subsection{{Computational Overhead}}
\label{sec:computational_overhead}

{Table~\ref{tab:computational} summarizes the parameter counts, GFLOPs, and per-batch wall-clock times for all methods, measured on an NVIDIA H100 80\,GB GPU with batch size 64 and input shape $(64 \times 3000 \times 270)$ using 10 warm-up and 50 timed iterations. Since SFUDA assumes access only to a pretrained source model and unlabeled target data, we report adaptation and inference costs only.}

{SHOT-IM relies solely on entropy minimization during adaptation, contributing 107.71\,GFLOPs per batch. SHOT++ and MU-SHOT-Fi additionally perform rotation SSL pre-training (215.43\,GFLOPs) before the joint adaptation loop, resulting in 323.14\,GFLOPs per batch during adaptation. The SSL pre-training stage (0.009\,s per batch) is run only once prior to adaptation and does not contribute to per-epoch adaptation time. Despite this additional pre-training cost, the wall-clock time of MU-SHOT-Fi during adaptation (0.234\,s per batch) remains comparable to SHOT++ (0.232\,s per batch), as both share an identical forward pass structure.}

{After deployment, all methods use only the main backbone components ($F_\eta$, $B_\psi$, $C_\phi$) for inference, incurring 107.71\,GFLOPs per batch and a wall-clock time of 0.003\,s per batch (0.058\,ms per sample). This decoupling enables a practical deployment strategy: adaptation is performed once on a server with sufficient resources, after which only the lightweight adapted model is deployed at the edge. The total parameter count of MU-SHOT-Fi is 910,976 during adaptation (886,210 for $F_\eta$, 16,512 for $B_\psi$, 7,740 for $C_\phi$, and 514 for the rotation head $R_\gamma$), reducing to 910,462 at inference once $R_\gamma$ is discarded.}

\begin{table}[t]
\centering
\caption{{Computational overhead per batch (batch size = 64, input $64 \times 3000 \times 270$) on an NVIDIA H100 80\,GB GPU (10 warm-up / 50 timed iterations). $R_\gamma$ denotes the rotation head used only during adaptation.}}
\label{tab:computational}
{
\resizebox{\columnwidth}{!}{
\begin{tabular}{llccc}
\toprule
Method & Stage & Parameters & GFLOPs & Time/Batch (s) \\
\midrule
SHOT-IM & Adaptation (entropy min.) & 910,462 & 107.71 & 0.1035 \\
\midrule
\multirow{2}{*}{SHOT++} & SSL pre-training of $R_\gamma$ (once) & 903,236 & 215.43 & 0.0087 \\
 & Joint adaptation & 910,976 & 323.14 & 0.2313 \\
\midrule
\multirow{2}{*}{MU-SHOT-Fi} & SSL pre-training of $R_\gamma$ (once) & 903,236 & 215.43 & 0.0087 \\
 & Joint adaptation & 910,976 & 323.14 & 0.2337 \\
\midrule
All methods & Inference ($F_\eta + B_\psi + C_\phi$) & 910,462 & 107.71 & 0.0037 \\
\bottomrule
\end{tabular}
}
}
\end{table}

\section{Numerical Results and Discussions}
\label{sec:results}

\subsection{MU-SHOT-Fi Results and Analysis}

Table~\ref{tab:wimans_vertical} presents comprehensive performance metrics for MU-SHOT-Fi across three domain shift scenarios on the WiMANS dataset. Our experimental results reveal a consistent pattern: adaptation gains scale inversely with source-only baseline performance. This relationship reflects a fundamental principle in domain adaptation: when the initial domain gap is large (weak source-only performance), there exists greater headroom for improvement through adaptation.
Specifically, scenarios exhibiting strong domain mismatch (Cross-Frequency and Combined shifts) demonstrate near-complete source model failure (0.00\% Exact Match, approximately 20\% Slot-wise Accuracy), yet achieve substantial recovery through adaptation. In contrast, the Cross-Room scenario begins from a relatively stronger baseline (55.09\% Slot-wise Accuracy), with limited improvements due to the reduced domain mismatch.

{All results are reported as mean $\pm$ std across 3 independent runs with different random seeds. The non-overlapping standard deviation intervals between MU-SHOT-Fi and source-only baselines confirm the reliability of the reported improvements. For example, under Combined Shift, MU-SHOT-Fi achieves $41.97\% \pm 2.39$ Slot-wise Accuracy compared to $19.61\% \pm 4.08$ for source-only.}

{We note that Exact Match is an inherently strict metric due to the combinatorial output space: with $M = 6$ slots each over $K+1 = 10$ classes, the joint label space contains $10^6 = 1{,}000{,}000$ configurations. A uniformly random predictor achieves Exact Match of approximately $(1/10)^6 \approx 0.0001\%$. In comparison, MU-SHOT-Fi attains $2.88\%$ average Exact Match, representing an improvement of approximately $28{,}800\times$ over random chance. To the best of our knowledge, prior works on WiMANS and related multi-user sensing benchmarks do not report Exact Match even under full supervision~\cite{tan2019multitrack, huang2024wimans, rizk2025multisensex}, reflecting a broader consensus that slot-level accuracy and occupancy estimation provide more stable measures of multi-user performance. We include Exact Match as an additional stringent evaluation perspective and report a random predictor baseline in Table~\ref{tab:wimans_vertical} for reference.}

\subsubsection{Cross-Room Domain Shift}
Source-only models attain 55.09\% slot-wise Accuracy but only 2.12\% Exact Match, indicating that while individual slots are often correct, full multi-user configurations are rarely recovered. MU-SHOT-Fi increases Exact Match to 7.76\% while maintaining SlotAcc (55.29\%) and achieving best occupancy performance (OccMAE 1.60 vs. 1.52 source-only; OccExact 24.61\%). 


\subsubsection{Cross-Frequency Domain Shift}
Frequency-dependent propagation effects induce severe distribution shift. Source-only models completely fail at sample level (0.00\% Exact Match, 21.16\% Slot-wise Accuracy). Both MU-SHOT-Fi and SHOT++ recover to 0.71\% Exact Match and ~47\% Slot-wise Accuracy with identical OccMAE (1.89), representing a 26-percentage-point improvement over source-only baseline. This recovery demonstrates that occupancy-weighted information maximization combined with rotation SSL effectively realigns features under frequency-induced domain shift.

\subsubsection{Combined Environment and Frequency Shift}
The combined room-and-frequency shift induces the most severe domain mismatch, yielding near-zero Exact Match for all methods and weak source-only generalization (19.61\% Slot-wise Accuracy). MU-SHOT-Fi achieves best Slot-wise Accuracy (41.97\%), lowest Occupancy MAE (2.12 vs. 3.00 source-only), and highest Occupancy Exact Match (10.93\%). While MU-SHOT-Fi ties SHOT++ in Exact Match (0.18\%), its consistent gains across slot correctness and occupancy metrics indicate that occupancy-aware information maximization yields more stable adaptation under compound domain shifts.

Notably, source-only models in cross-frequency and combined-shift settings attain relatively high ActivityF1 (25.69\%, 24.99\%) despite 0\% Exact Match. This discrepancy arises because ActivityF1 is computed only over occupied ground-truth slots and does not penalize errors in occupancy counting, extra predicted users, or inconsistent slot assignment after matching. Exact Match and SlotAcc instead reflect the full end-task requirement-correct counting and coherent multi-user slot configuration-so improvements in occupancy consistency substantially increase these metrics without proportional gains in ActivityF1.

\begin{table*}[t]
\centering
\caption{WiMANS set-prediction head ablation: pooled parallel slot classifier vs.\ query-based transformer decoder (DETR-style). Both variants use the same SFUDA pipeline (Hungarian matching in source training; occupancy-weighted GENT and rotation-based self-supervised learning during target adaptation). Values are mean$\pm$std across runs. Best and second-best within each scenario are \textbf{bolded} and \underline{underlined}, respectively.}
\label{tab:detr_ablation}
\scriptsize
\centering
\begin{tabular}{>{\centering\arraybackslash}p{2.0cm}
                >{\centering\arraybackslash}p{4.0cm}
                >{\centering\arraybackslash}p{2.0cm}
                >{\centering\arraybackslash}p{2.0cm}
                >{\centering\arraybackslash}p{2.0cm}
                >{\centering\arraybackslash}p{2.0cm}
                >{\centering\arraybackslash}p{1.0cm}}
\toprule
\textbf{Scenario} & \textbf{Architecture} & \textbf{\shortstack{Slot-wise\\Acc (\%) $\uparrow$}} & \textbf{\shortstack{Activity\\F1 (\%) $\uparrow$}} & \textbf{\shortstack{Exact\\Match (\%) $\uparrow$}} & \textbf{\shortstack{Occ.\\MAE $\downarrow$}} & \textbf{\shortstack{$\Delta$ Slot-wise\\Acc (\%)}} \\
\midrule
\multirow{4}{*}{Cross-Room}
& Source-only (Pooled Parallel)          & 55.09 $\pm$ 2.47 & 15.89 $\pm$ 0.47 & 2.12 $\pm$ 0.43 & \underline{1.52 $\pm$ 0.24} & - \\
& MU-SHOT-Fi (Pooled Parallel)           & 55.29 $\pm$ 1.98 & \textbf{17.08 $\pm$ 1.15} & \underline{7.76 $\pm$ 1.51} & 1.60 $\pm$ 0.25 & \textbf{+0.20} \\
& Source-only (Query Decoder)            & \textbf{58.55 $\pm$ 0.25} & 12.45 $\pm$ 0.50 & \textbf{8.64 $\pm$ 1.31} & 1.53 $\pm$ 0.07 & - \\
& MU-SHOT-Fi (Query Decoder)             & \underline{58.00 $\pm$ 1.06} & \underline{16.07 $\pm$ 0.36} & 7.59 $\pm$ 0.99 & \textbf{1.48 $\pm$ 0.15} & -0.55 \\
\midrule
\multirow{4}{*}{\shortstack[l]{Cross-\\Frequency}}
& Source-only (Pooled Parallel)          & 21.16 $\pm$ 4.71 & \textbf{25.69 $\pm$ 0.97} & 0.00 $\pm$ 0.00 & 2.92 $\pm$ 0.33 & - \\
& MU-SHOT-Fi (Pooled Parallel)           & 46.97 $\pm$ 1.35 & 18.82 $\pm$ 1.50 & 0.71 $\pm$ 0.25 & 1.89 $\pm$ 0.15 & \textbf{+25.81} \\
& Source-only (Query Decoder)            & \textbf{55.87 $\pm$ 4.26} & 8.34 $\pm$ 1.52 & \textbf{3.35 $\pm$ 2.81} & \underline{1.74 $\pm$ 0.25} & - \\
& MU-SHOT-Fi (Query Decoder)             & \underline{54.02 $\pm$ 2.38} & \underline{19.26 $\pm$ 3.71} & \underline{1.41 $\pm$ 0.25} & \textbf{1.65 $\pm$ 0.24} & -1.85 \\
\midrule
\multirow{4}{*}{Combined}
& Source-only (Pooled Parallel)          & 19.61 $\pm$ 4.08 & \textbf{24.99 $\pm$ 2.53} & 0.00 $\pm$ 0.00 & 3.00 $\pm$ 0.35 & - \\
& MU-SHOT-Fi (Pooled Parallel)           & 41.97 $\pm$ 2.39 & \underline{18.47 $\pm$ 2.91} & 0.18 $\pm$ 0.25 & 2.12 $\pm$ 0.12 & \textbf{+22.36} \\
& Source-only (Query Decoder)            & \textbf{54.26 $\pm$ 2.49} & 4.74 $\pm$ 0.88 & \underline{0.35 $\pm$ 0.25} & \underline{1.94 $\pm$ 0.99} & - \\
& MU-SHOT-Fi (Query Decoder)             & \underline{53.35 $\pm$ 1.75} & 13.48 $\pm$ 2.36 & \textbf{1.41 $\pm$ 0.25} & \textbf{1.66 $\pm$ 0.07} & -0.91 \\
\midrule
\multirow{4}{*}{Average}
& Source-only (Pooled Parallel)          & 31.95 $\pm$ 3.75 & \textbf{22.19 $\pm$ 1.32} & 0.71 $\pm$ 0.14 & 2.48 $\pm$ 0.31 & - \\
& MU-SHOT-Fi (Pooled Parallel)           & 48.08 $\pm$ 1.91 & \underline{18.12 $\pm$ 1.85} & 2.88 $\pm$ 0.67 & 1.87 $\pm$ 0.17 & \textbf{+16.13} \\
& Source-only (Query Decoder)            & \textbf{56.23 $\pm$ 2.33} & 8.51 $\pm$ 0.97 & \textbf{4.11 $\pm$ 1.46} & \underline{1.74 $\pm$ 0.44} & - \\
& MU-SHOT-Fi (Query Decoder)             & \underline{55.12 $\pm$ 1.73} & 16.27 $\pm$ 2.14 & \underline{3.47 $\pm$ 0.50} & \textbf{1.60 $\pm$ 0.15}  & -1.11 \\
\bottomrule
\vspace{1mm}
\end{tabular}

\raggedright
\footnotesize
Note: $\Delta$ Slot-wise Acc.\ (\%) = (Adapted) $-$ (Source-only), computed using the mean Slot-wise Accuracy for each scenario. Adapted variants use occupancy-weighted GENT, Hungarian matching, and rotation-based self-supervised learning. The pooled parallel classifier uses global average pooling followed by per-slot linear classification. The query-based decoder preserves spatial tokens and predicts slots using a 6-layer transformer decoder with learned queries.
\end{table*}

\subsection{Ablation Studies}
\label{subsec:cpc_analysis}

\subsubsection{Integrating CPC within MU-SHOT-Fi}
Table~IV depicts that adding CPC to MU-SHOT-Fi provides no benefit and often hurts performance. In Cross-Room, Exact Match drops from 7.76\% to 6.34\%. In Combined shift, Slot-wise Acc decreases from 41.97\% to 41.58\%. Only Cross-Frequency shows marginal improvement (Exact Match: 0.71\% to 1.41\%), but this is offset by degradation elsewhere. The degradation of CPC in multi-user scenarios stems from a fundamental conflict: CPC enforces sample-level temporal consistency by predicting future windows from past context, but multi-user prediction requires permutation-invariant slot outputs. Hungarian matching reorders slots independently per sample, breaking the temporal coherence CPC attempts to learn. This mismatch causes CPC's gradients to conflict with the slot-prediction objective, degrading rather than improving adaptation. We thus exclude CPC from MU-SHOT-Fi, using only rotation SSL for spatial regularization combined with occupancy-weighted information maximization. This design achieves the best average performance across metrics (Table~\ref{tab:wimans_vertical}, bottom rows).

\begin{figure}[h]
\centering
\includegraphics[width=\columnwidth]{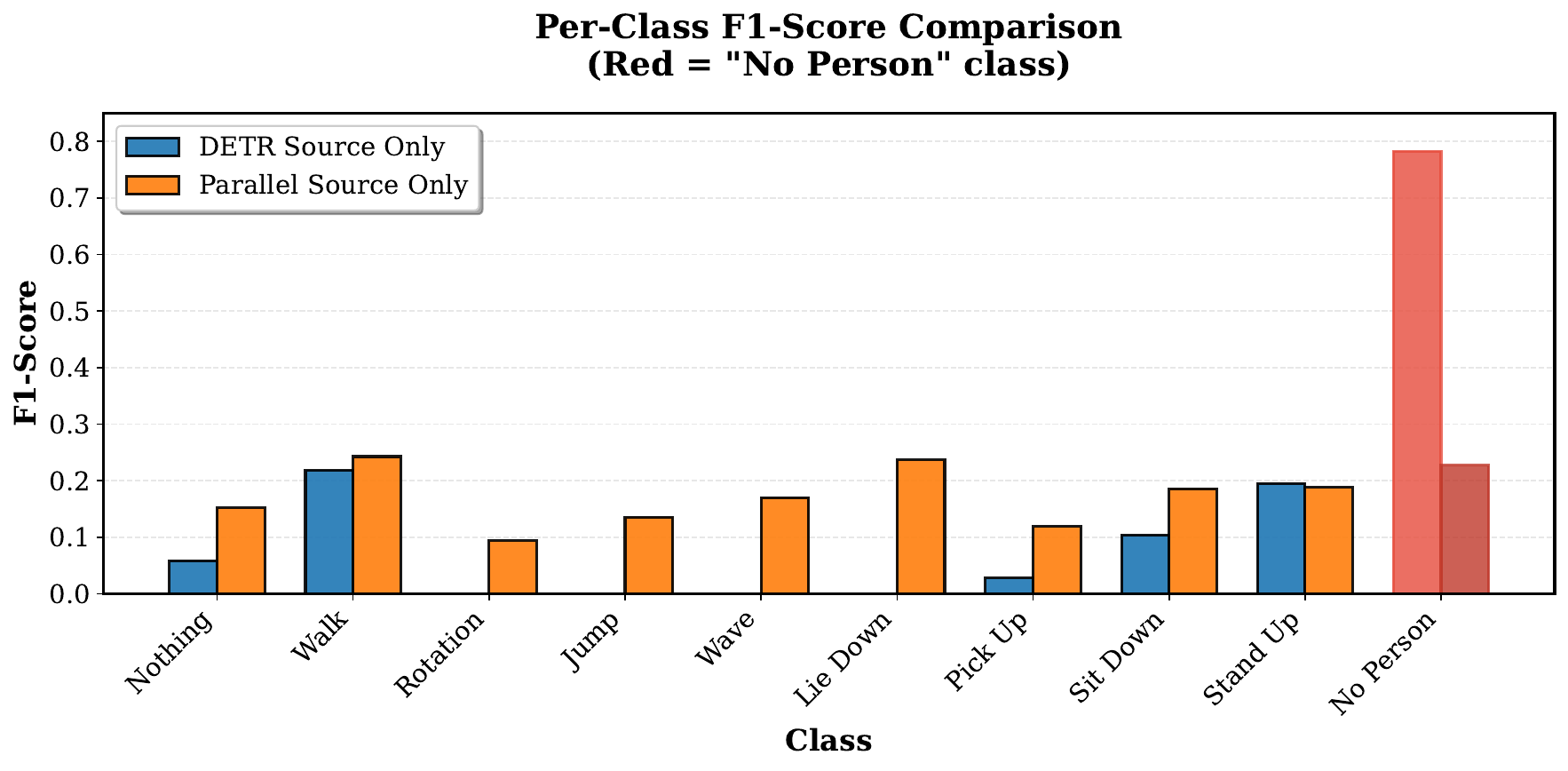}
\caption{Per-class F1-scores under Cross-Frequency shift (source-only). DETR achieves 77.1\% F1 on "no person" (red) but only 14.5\% average on activities, demonstrating majority-class exploitation. Parallel heads show more balanced performance (23.1\% vs 16.3\%).}
\label{fig:detr_f1}
\end{figure}

\subsubsection{Set-Prediction Head Ablation: Pooled Parallel vs.\ Query-Based Decoder}
\label{subsec:detr_ablation}

In Table~\ref{tab:detr_ablation}, we compare two architectures for multi-user slot prediction: (i) a pooled parallel classifier that averages CNN features globally then predicts each slot independently, and (ii) a query-based transformer decoder that preserves spatial structure and uses attention to predict slots~\cite{carion2020end}. Both use Hungarian matching during source training and occupancy-weighted GENT with rotation SSL during adaptation.
Table~\ref{tab:detr_ablation} shows the query-based decoder achieves higher source-only Slot-wise Accuracy (56.23\% vs.\ 31.95\%), but this comes from predicting ``no person'' rather than learning activities. That is, ActivityF1 is only 8.51\% compared to 22.19\% for the pooled classifier. Fig.~\ref{fig:detr_f1} confirms this under cross-frequency: the decoder attains 77.1\% F1 on ``no person'' but 14.5\% on activities, while the pooled classifier is balanced (23.1\% vs.\ 16.3\%).
During adaptation, the pooled classifier improves Slot-wise Accuracy by +16.13\%, while the decoder degrades ($-1.11\%$). This shows the decoder's majority-class bias harms adaptation. We adopt the pooled parallel architecture for lower complexity and better adaptation.

\subsubsection{{Hyperparameter Sensitivity}}
{We analyze the sensitivity of MU-SHOT-Fi to the self-supervised loss weight $\lambda_{\text{rot}}$ under Cross-Room shift, with $\lambda_{\text{ent}} = 1.0$ fixed. Table~\ref{tab:sensitivity} reports Exact Match and Occupancy Exact Match across $\lambda_{\text{rot}} \in \{0.0, 0.1, 0.5, 1.0\}$.}
{We note that performance remains stable across all nonzero configurations, with identical Exact Match (7.41\%) and Occupancy Exact Match (22.93\%) for $\lambda_{\text{rot}} \in \{0.1, 0.5, 1.0\}$. This stability indicates that adaptation is primarily constrained by the source representation quality and task difficulty, rather than the precise loss weighting - as long as both $\mathcal{L}_{\text{ent}}$ and $\mathcal{L}_{\text{rot}}$ contribute meaningful gradients toward the frozen classifier hypothesis, the backbone converges to the same solution, consistent with findings in SHOT++~\cite{liang2021source}. In contrast, setting $\lambda_{\text{rot}} = 0$ (equivalent to SHOT-IM without rotation SSL) causes Exact Match to drop to 2.82\% and Occupancy Exact Match to 17.99\%, with substantially higher variance across runs, confirming that rotation SSL is necessary for stable adaptation but that its precise weight does not affect the converged solution once active.}
\begin{table}[t]
\centering
\caption{Hyperparameter sensitivity of MU-SHOT-Fi to $\lambda_{\text{rot}}$ under Cross-Room shift ($\lambda_{\text{ent}} = 1.0$ fixed). Values are mean $\pm$ std across 3 runs. Higher is better ($\uparrow$).}
\label{tab:sensitivity}
{
\begin{tabular}{ccc}
\toprule
$\lambda_{\text{rot}}$ & Exact Match (\%) $\uparrow$ & Occ.\ Exact Match (\%) $\uparrow$ \\
\midrule
0.0 (SHOT-IM) & 2.82 $\pm$ 2.94 & 17.99 $\pm$ 7.68 \\
0.1 & 7.41 $\pm$ 2.41 & 22.93 $\pm$ 4.62 \\
0.5 (default) & 7.41 $\pm$ 2.41 & 22.93 $\pm$ 4.62 \\
1.0 & 7.41 $\pm$ 2.41 & 22.93 $\pm$ 4.62 \\
\bottomrule
\end{tabular}
}
\end{table}


\begin{table}[t]
\centering
\caption{Widar 3.0: Source-free domain adaptation results under three domain shifts. Values are mean $\pm$ std across runs. Higher is better ($\uparrow$). Best within each scenario is \textbf{bolded} and second-best is \underline{underlined}.}
\label{tab:widar_results}
\resizebox{\columnwidth}{!}{%
\begin{tabular}{@{}llcccc@{}}
\toprule
\textbf{Scenario} & \textbf{Method} & \textbf{Accuracy (\%) $\uparrow$} & \textbf{F1-Macro (\%) $\uparrow$} & \textbf{Gain (\%)} \\
\midrule
\multirow{4}{*}{Cross-Room} 
& Source Only           & 69.09 $\pm$ 0.57 & 68.37 $\pm$ 0.58 & - \\
& SHOT-IM               & 78.19 $\pm$ 1.13 & 77.08 $\pm$ 0.89 & +9.10 \\
& \quad + SSL (SHOT++)  & \underline{79.16 $\pm$ 1.88} & \underline{77.97 $\pm$ 1.54} & +10.07 \\
& \quad + CPC (SU-SHOT-Fi)    & \textbf{79.91 $\pm$ 1.72} & \textbf{78.67 $\pm$ 1.21} & \textbf{+10.82} \\
\midrule
\multirow{4}{*}{Cross-Torso}
& Source Only           & 87.73 $\pm$ 2.28 & 88.15 $\pm$ 1.83 & - \\
& SHOT-IM               & \underline{89.07 $\pm$ 1.75} & \underline{89.61 $\pm$ 1.42} & +1.34 \\
& \quad + SSL (SHOT++)  & \textbf{89.69 $\pm$ 1.74} & \textbf{90.25 $\pm$ 1.41} & +1.96 \\
& \quad + CPC (SU-SHOT-Fi)    & \textbf{89.69 $\pm$ 2.00} & \textbf{90.25 $\pm$ 1.58} & +1.96 \\
\midrule
\multirow{4}{*}{Cross-Face}
& Source Only           & 83.73 $\pm$ 2.82 & 84.41 $\pm$ 2.48 & - \\
& SHOT-IM               & \underline{87.56 $\pm$ 1.08} & \underline{88.27 $\pm$ 0.90} & +3.83 \\
& \quad + SSL (SHOT++)  & \textbf{87.64 $\pm$ 1.11} & \textbf{88.39 $\pm$ 0.91} & +3.91 \\
& \quad + CPC (SU-SHOT-Fi)    & \textbf{87.64 $\pm$ 1.32} & 88.26 $\pm$ 1.01 & +3.91 \\
\midrule
\multirow{4}{*}{Average}
& Source Only           & 80.18 $\pm$ 1.89 & 80.31 $\pm$ 1.63 & - \\
& SHOT-IM               & 84.94 $\pm$ 1.32 & 84.99 $\pm$ 1.07 & +4.76 \\
& \quad + SSL (SHOT++)  & \underline{85.50 $\pm$ 1.58} & \underline{85.54 $\pm$ 1.29} & +5.32 \\
& \quad + CPC (SU-SHOT-Fi)    & \textbf{85.75 $\pm$ 1.68} & \textbf{85.73 $\pm$ 1.27} & \textbf{+5.57} \\
\bottomrule
\vspace{1mm}
\end{tabular}
}

\raggedright
\footnotesize
Note: Average row shows mean performance across the three domain shift scenarios. Per-class F1-scores are detailed in Table~\ref{tab:widar_perclass}.
\end{table}

\subsection{SU-SHOT-Fi Results and Analysis}
\label{subsec:widar_results}
Table~\ref{tab:widar_results} reports classification accuracy across three domain shift scenarios. SU-SHOT-Fi achieves best or tied-best performance in all settings, with gains scaling inversely with source-only baseline strength. Source baselines reflect varying domain shift severity: Cross-Room (69.09\%), Cross-Face (83.73\%), Cross-Torso (87.73\%). Adaptation gains correspondingly decrease from +10.82\% (Cross-Room) to +3.91\% (Cross-Face) to +1.96\% (Cross-Torso). This inverse scaling aligns with domain adaptation theory~\cite{ben2010theory}: when source and target distributions exhibit low divergence, source features already generalize well, constraining adaptation headroom. 

SU-SHOT-Fi's advantage over SHOT++ is largest in Cross-Room (+0.75\%), where CPC-based temporal modeling exploits domain-invariant periodic patterns (e.g., stride rhythms) that remain stable despite environmental layout changes. In Cross-Torso and Cross-Face, where source baselines exceed 83\%, all adaptation methods (SHOT, SHOT++, SU-SHOT-Fi) converge to similar performance as the adaptation ceiling is approached. Averaged across scenarios, SU-SHOT-Fi improves over SHOT-IM by 0.81\% and over source-only by 5.57\%.

Table~\ref{tab:widar_perclass} reports per-class F1-scores. SU-SHOT-Fi achieves best average performance on 5 of 6 gestures. Performance heterogeneity reveals underlying CSI signature characteristics: \emph{Slide} (74.11\% F1) produces low Doppler content and gradual amplitude variations that blend with background noise under domain shift, while \emph{Draw-Zigzag(H)} (92.11\% F1) generates pronounced multipath variations from rapid directional reversals, creating robust discriminative signatures.

SU-SHOT-Fi yields largest gains on temporal-rich gestures: \emph{Clap} improves by 7.78\% (81.00\% → 88.78\%) and \emph{Sweep} by 4.78\%. This pattern validates CPC's temporal modeling: clapping produces periodic amplitude bursts with consistent inter-burst intervals that CPC captures through window-level prediction, while sweeping creates continuous Doppler shifts spanning multiple temporal windows. In contrast, gestures with static phases interspersed with motion-\emph{Push\&Pull} (+2.66\%) and \emph{Draw-O(H)} (+5.34\%)-benefit less from CPC, as temporal coherence is disrupted by motion-pause transitions where rotation SSL's spatial regularization dominates.


\begin{table}[t]
\centering
\caption{Widar 3.0: Per-class F1-scores (\%) across three domain shifts. Results show average performance over 3 runs. Best within each scenario is \textbf{bolded} and second-best is \underline{underlined}.}
\label{tab:widar_perclass}
\small
\setlength{\tabcolsep}{3pt}
\resizebox{\columnwidth}{!}{%
\begin{tabular}{@{}llcccc@{}}
\toprule
\textbf{Gesture} & \textbf{Method} & \textbf{Cross-Room} & \textbf{Cross-Torso} & \textbf{Cross-Face} & \textbf{Average} \\
\midrule
\multirow{4}{*}{Clap}
& Source Only              & 72.67 & 83.00 & 87.33 & 81.00 \\
& SHOT-IM                  & 84.00 & 86.67 & \textbf{91.00} & 87.22 \\
& \quad + SSL (SHOT++)     & \underline{86.33} & \underline{87.33} & \underline{90.00} & \underline{87.89} \\
& \quad + CPC (SU-SHOT-Fi)       & \textbf{87.67} & \textbf{87.67} & \textbf{91.00} & \textbf{88.78} \\
\midrule
\multirow{4}{*}{Draw-O(H)}
& Source Only              & 55.67 & 88.00 & 91.33 & 78.33 \\
& SHOT-IM                  & 67.67 & 88.67 & \textbf{92.33} & 82.89 \\
& \quad + SSL (SHOT++)     & \underline{68.67} & \underline{89.67} & \underline{91.67} & \underline{83.34} \\
& \quad + CPC (SU-SHOT-Fi)       & \textbf{69.67} & \textbf{90.33} & 91.00 & \textbf{83.67} \\
\midrule
\multirow{4}{*}{\shortstack[l]{Draw-\\Zigzag(H)}}
& Source Only              & 71.67 & \underline{97.33} & 93.00 & 87.33 \\
& SHOT-IM                  & \underline{82.33} & \textbf{98.33} & \textbf{95.00} & \underline{91.89} \\
& \quad + SSL (SHOT++)     & 82.00 & \textbf{98.33} & \textbf{95.00} & 91.78 \\
& \quad + CPC (SU-SHOT-Fi)       & \textbf{83.67} & \textbf{98.33} & \underline{94.33} & \textbf{92.11} \\
\midrule
\multirow{4}{*}{Push\&Pull}
& Source Only              & 88.67 & 90.67 & 89.33 & 89.56 \\
& SHOT-IM                  & \underline{90.33} & 91.67 & 93.00 & 91.67 \\
& \quad + SSL (SHOT++)     & \textbf{91.33} & \textbf{92.67} & \textbf{94.00} & \textbf{92.67} \\
& \quad + CPC (SU-SHOT-Fi)       & \textbf{91.33} & \underline{92.00} & \underline{93.33} & \underline{92.22} \\
\midrule
\multirow{4}{*}{Slide}
& Source Only              & 40.67 & \underline{83.33} & 74.00 & 66.00 \\
& SHOT-IM                  & 57.33 & 82.67 & 79.67 & 73.22 \\
& \quad + SSL (SHOT++)     & \underline{58.00} & \textbf{83.67} & \underline{80.00} & \underline{73.89} \\
& \quad + CPC (SU-SHOT-Fi)       & \textbf{58.33} & \underline{83.33} & \textbf{80.67} & \textbf{74.11} \\
\midrule
\multirow{4}{*}{Sweep}
& Source Only              & 77.33 & 86.00 & 71.67 & 78.33 \\
& SHOT-IM                  & 76.00 & 89.00 & \underline{79.33} & 81.44 \\
& \quad + SSL (SHOT++)     & \underline{80.67} & \textbf{90.00} & \textbf{79.67} & \textbf{83.44} \\
& \quad + CPC (SU-SHOT-Fi)       & \textbf{81.00} & \underline{89.67} & 78.67 & \underline{83.11} \\
\midrule
\multirow{4}{*}{Macro Avg}
& Source Only              & 67.78 & 88.06 & 84.44 & 80.09 \\
& SHOT-IM                  & 76.28 & 89.50 & \textbf{88.39} & 84.72 \\
& \quad + SSL (SHOT++)     & \underline{77.83} & \textbf{90.28} & \textbf{88.39} & \underline{85.50} \\
& \quad + CPC (SU-SHOT-Fi)       & \textbf{78.61} & \underline{90.22} & \underline{88.17} & \textbf{85.67} \\
\bottomrule
\end{tabular}
}
\vspace{1mm}

\raggedright
\footnotesize
Note: Macro Avg computed across the six gesture classes. Average column shows mean F1-scores across the three domain shift scenarios.
\end{table}

\section{Conclusion}
\label{sec:conclusion}

We introduced MU-SHOT-Fi, the first SFUDA framework for multi-user Wi-Fi sensing. Multi-user scenarios present fundamental challenges such as variable occupancy, severe class imbalance, and signal entanglement from concurrent activities. Standard pseudo-labeling collapses under this imbalance, converging to predict ``no person'' while ignoring actual activities. Our key contribution, occupancy-weighted information maximization, prevents this collapse by down-weighting empty slots during diversity regularization, enabling adaptation across variable user counts without explicit occupancy estimation. Binary rotation SSL provides spatial regularization on CSI's frequency-time structure. 
Evaluation on WiMANS demonstrates adaptation gains scaling inversely with domain shift severity. Notably, our ablation reveals that CPC, while effective for single-user scenarios, degrades multi-user performance due to conflicts between sample-level temporal consistency and permutation-invariant slot predictions.

\bibliographystyle{IEEEtran}
\bibliography{references}

@article{radwan2025tutorial,
  title={A Tutorial-cum-Survey on Self-Supervised Learning for Wi-Fi Sensing: Trends, Challenges, and Outlook},
  author={Radwan, Ahmed Y and Yildirim, Mustafa and Hasanzadeh, Navid and Tabassum, Hina and Valaee, Shahrokh},
  journal={IEEE Commun. Surveys \& Tut.},
  year={2025},
  publisher={IEEE}
}

@inproceedings{hasanzadeh2024enhancing,
  title={Enhancing generalization in human activity recognition through improved Wi-Fi channel state information phase processing and antenna pair selection},
  author={Hasanzadeh, Navid and Valaee, Shahrokh},
  booktitle={2024 IEEE 34th International Workshop on Machine Learning for Signal Processing (MLSP)},
  pages={1--6},
  year={2024},
  organization={IEEE}
}

@article{wu2022wifi,
  title={WiFi CSI-based device-free sensing: from Fresnel zone model to CSI-ratio model},
  author={Wu, Dan and Zeng, Youwei and Zhang, Fusang and Zhang, Daqing},
  journal={CCF Trans. on Pervasive Computing and Interaction},
  volume={4},
  number={1},
  pages={88--102},
  year={2022},
  publisher={Springer}
}

@article{wei2025survey,
  title={A survey on WiFi-based human identification: Scenarios, challenges, and current solutions},
  author={Wei, Zhongcheng and Chen, Wei and Ning, Shuli and Lin, Weidong and Li, Nan and Lian, Bin and Sun, Xiang and Zhao, Jijun},
  journal={ACM Trans. on Sensor Networks},
  volume={21},
  number={1},
  pages={1--32},
  year={2025},
  publisher={ACM New York, NY}
}

@article{luo2018channel,
  title={Channel state information prediction for 5G wireless communications: A deep learning approach},
  author={Luo, Changqing and Ji, Jinlong and Wang, Qianlong and Chen, Xuhui and Li, Pan},
  journal={IEEE Trans. on network science and engineering},
  volume={7},
  number={1},
  pages={227--236},
  year={2018},
  publisher={IEEE}
}

@article{berthelot2019mixmatch,
  title={Mixmatch: A holistic approach to semi-supervised learning},
  author={Berthelot, David and Carlini, Nicholas and Goodfellow, Ian and Papernot, Nicolas and Oliver, Avital and Raffel, Colin A},
  journal={Advances in neural information processing Sys.},
  volume={32},
  year={2019}
}

@article{liang2021source,
  title={Source data-absent unsupervised domain adaptation through hypothesis transfer and labeling transfer},
  author={Liang, Jian and Hu, Dapeng and Wang, Yunbo and He, Ran and Feng, Jiashi},
  journal={IEEE Trans. on Pattern Analysis and Machine Intelligence},
  volume={44},
  number={11},
  pages={8602--8617},
  year={2021},
  publisher={IEEE}
}

@inproceedings{liang2020we,
  title={Do we really need to access the source data? source hypothesis transfer for unsupervised domain adaptation},
  author={Liang, Jian and Hu, Dapeng and Feng, Jiashi},
  booktitle={International Conf. on machine learning},
  pages={6028--6039},
  year={2020},
  organization={PMLR}
}

@article{ma2019wifisensing,
  author  = {Ma, Yongsen and Zhou, Gang and Wang, Shuangquan and Zhao, Hongyang and Jung, Woosub},
  title   = {WiFi Sensing with Channel State Information: A Survey},
  journal = {ACM Computing Surveys},
  volume  = {52},
  number  = {3},
  pages   = {46:1--46:36},
  year    = {2019},
  doi     = {10.1145/3310194}
}

@article{zhang2021widar3,
  author  = {Zhang, Yi and Zheng, Yue and Qian, Kun and Zhang, Guidong and Liu, Yunhao and Wu, Chenshu and Yang, Zheng},
  title   = {Widar3.0: Zero-Effort Cross-Domain Gesture Recognition With Wi-Fi},
  journal = {IEEE Trans. on Pattern Analysis and Machine Intelligence},
  volume  = {44},
  number  = {11},
  pages   = {8671--8688},
  year    = {2022},
  doi     = {10.1109/TPAMI.2021.3051203}
}

@article{oord2018representation,
  title={Representation learning with contrastive predictive coding},
  author={Oord, Aaron van den and Li, Yazhe and Vinyals, Oriol},
  journal={arXiv preprint arXiv:1807.03748},
  year={2018}
}

@article{ben2010theory,
  title={A theory of learning from different domains},
  author={Ben-David, Shai and Blitzer, John and Crammer, Koby and Kulesza, Alex and Pereira, Fernando and Vaughan, Jennifer Wortman},
  journal={Machine learning},
  volume={79},
  number={1-2},
  pages={151--175},
  year={2010},
  publisher={Springer}
}

@inproceedings{huang2024wimans,
  title={Wimans: A benchmark dataset for wifi-based multi-user activity sensing},
  author={Huang, Shuokang and Li, Kaihan and You, Di and Chen, Yichong and Lin, Arvin and Liu, Siying and Li, Xiaohui and McCann, Julie A},
  booktitle={European Conf. on Computer Vision},
  pages={72--91},
  year={2024},
  organization={Springer}
}

@article{yan2025wi,
  title={Wi-SFDAGR: WiFi-Based Cross-Domain Gesture Recognition via Source-Free Domain Adaptation},
  author={Yan, Huan and Zhang, Xiang and Huang, Jinyang and Feng, Yuanhao and Li, Meng and Wang, Anzhi and Ou, Weihua and Wang, Hongbing and Liu, Zhi},
  journal={IEEE Internet of Things Journal},
  year={2025},
  publisher={IEEE}
}

@inproceedings{shi2020towards,
  title={Towards environment-independent behavior-based user authentication using wifi},
  author={Shi, Cong and Liu, Jian and Borodinov, Nick and Leao, Bruno and Chen, Yingying},
  booktitle={2020 IEEE 17th International Conf. on Mobile Ad Hoc and Sensor Sys. (MASS)},
  pages={666--674},
  year={2020},
  organization={IEEE}
}

@article{zhang2023unsupervised,
  title={Unsupervised domain adaptation for rf-based gesture recognition},
  author={Zhang, Bin-Bin and Zhang, Dongheng and Li, Yadong and Hu, Yang and Chen, Yan},
  journal={IEEE Internet of Things Journal},
  volume={10},
  number={23},
  pages={21026--21038},
  year={2023},
  publisher={IEEE}
}

@inproceedings{li2021deep,
  title={Deep transfer learning for {WiFi} localization},
  author={Li, Peizheng and Cui, Han and Khan, Aftab and Raza, Usman and Piechocki, Robert and Doufexi, Angela and Farnham, Tim},
  booktitle={2021 IEEE Radar Conf. (RadarConf21)},
  pages={1--5},
  year={2021},
  organization={IEEE}
}

@article{shi2022environment,
  title={{Environment-Robust {WiFi}-Based Human Activity Recognition Using Enhanced {CSI} and Deep Learning}},
  author={Shi, Zhenguo and Cheng, Qingqing and Zhang, J Andrew and Da Xu, Richard Yi},
  journal={IEEE Internet of Things Jrnl.},
  volume={9},
  number={24},
  pages={24643--24654},
  year={2022},
  publisher={IEEE}
}

@inproceedings{shirakami2021heart,
  title={Heart rate variability extraction using commodity {Wi-Fi} devices via time domain signal processing},
  author={Shirakami, Itsuki and Sato, Takashi},
  booktitle={2021 IEEE EMBS Intl. Conf. on Biomedical and Health Informatics (BHI)},
  pages={1--4},
  year={2021},
  organization={IEEE}
}

@inproceedings{raja2018wibot,
  title={WiBot! In-vehicle behaviour and gesture recognition using wireless network edge},
  author={Raja, Muneeba and Ghaderi, Viviane and Sigg, Stephan},
  booktitle={2018 IEEE 38th International Conf. on Distributed Computing Sys. (ICDCS)},
  pages={376--387},
  year={2018},
  organization={IEEE}
}

@article{ahmed2018estimating,
  title={Estimating angle-of-arrival and time-of-flight for multipath components using wifi channel state information},
  author={Ahmed, Afaz Uddin and Arablouei, Reza and De Hoog, Frank and Kusy, Branislav and Jurdak, Raja and Bergmann, Neil},
  journal={Sensors},
  volume={18},
  number={6},
  pages={1753},
  year={2018},
  publisher={MDPI}
}

@article{brunello2025time,
  title={Time matters: Empirical insights into the limits and challenges of temporal generalization in CSI-based Wi-Fi sensing},
  author={Brunello, Andrea and Montanari, Angelo and Montoliu, Ra{\'u}l and Moreira, Adriano and Saccomanno, Nicola and Sansano-Sansano, Emilio and Torres-Sospedra, Joaqu{\'\i}n},
  journal={Internet of Things},
  pages={101634},
  year={2025},
  publisher={Elsevier}
}

@article{chen2018wifi,
  title={WiFi CSI based passive human activity recognition using attention based BLSTM},
  author={Chen, Zhenghua and Zhang, Le and Jiang, Chaoyang and Cao, Zhiguang and Cui, Wei},
  journal={IEEE Trans. on Mobile Computing},
  volume={18},
  number={11},
  pages={2714--2724},
  year={2018},
  publisher={IEEE}
}

@article{wang2017tensorbeat,
  title={TensorBeat: Tensor decomposition for monitoring multiperson breathing beats with commodity WiFi},
  author={Wang, Xuyu and Yang, Chao and Mao, Shiwen},
  journal={ACM Trans. on Intelligent Sys. and Technology (TIST)},
  volume={9},
  number={1},
  pages={1--27},
  year={2017},
  publisher={ACM New York, NY, USA}
}

@article{rizk2025multisensex,
  title={MultiSenseX: A Sustainable Solution for Multi-Human Activity Recognition and Localization in Smart Environments},
  author={Rizk, Hamada and Elmogy, Ahmed and Rihan, Mohamed and Yamaguchi, Hirozumi},
  journal={AI},
  volume={6},
  number={1},
  pages={6},
  year={2025},
  publisher={MDPI}
}

@inproceedings{strohmayer2024data,
  title={Data augmentation techniques for cross-domain WiFi CSI-based human activity recognition},
  author={Strohmayer, Julian and Kampel, Martin},
  booktitle={IFIP International Conf. on Artificial Intelligence Applications and Innovations},
  pages={42--56},
  year={2024},
  organization={Springer}
}

@inproceedings{tan2019multitrack,
  title={MultiTrack: Multi-user tracking and activity recognition using commodity WiFi},
  author={Tan, Sheng and Zhang, Linghan and Wang, Zi and Yang, Jie},
  booktitle={Proceedings of the 2019 CHI Conf. on Human Factors in Computing Sys.},
  pages={1--12},
  year={2019}
}

@article{yousefi2017survey,
  title={A survey on behavior recognition using WiFi channel state information},
  author={Yousefi, Siamak and Narui, Hirokazu and Dayal, Sankalp and Ermon, Stefano and Valaee, Shahrokh},
  journal={IEEE Commun. Magazine},
  volume={55},
  number={10},
  pages={98--104},
  year={2017},
  publisher={IEEE}
}

@inproceedings{cominelli2023exposing,
  title={Exposing the csi: A systematic investigation of csi-based wi-fi sensing capabilities and limitations},
  author={Cominelli, Marco and Gringoli, Francesco and Restuccia, Francesco},
  booktitle={2023 IEEE International Conf. on Pervasive Computing and Commun. (PerCom)},
  pages={81--90},
  year={2023},
  organization={IEEE}
}

@article{ding2024multiple,
  title={A Multiple WiFi Sensors Assisted Human Activity Recognition Scheme for Smart Home},
  author={Ding, Jianyang and Wang, Yong and Xie, Qian and Niu, Jiajun},
  journal={IEEE Sensors Journal},
  year={2024},
  publisher={IEEE}
}

@article{chen2022fidora,
  title={Fidora: Robust WiFi-based indoor localization via unsupervised domain adaptation},
  author={Chen, Xi and Li, Hang and Zhou, Chenyi and Liu, Xue and Wu, Di and Dudek, Gregory},
  journal={IEEE Internet of Things Journal},
  volume={9},
  number={12},
  pages={9872--9888},
  year={2022},
  publisher={IEEE}
}

@article{gidaris2018unsupervised,
  title={Unsupervised representation learning by predicting image rotations},
  author={Gidaris, Spyros and Singh, Praveer and Komodakis, Nikos},
  journal={arXiv preprint arXiv:1803.07728},
  year={2018}
}

@article{grill2020bootstrap,
  title={Bootstrap your own latent-a new approach to self-supervised learning},
  author={Grill, Jean-Bastien and Strub, Florian and Altch{\'e}, Florent and Tallec, Corentin and Richemond, Pierre and Buchatskaya, Elena and Doersch, Carl and Avila Pires, Bernardo and Guo, Zhaohan and Gheshlaghi Azar, Mohammad and others},
  journal={Advances in neural information processing Sys.},
  volume={33},
  pages={21271--21284},
  year={2020}
}

@article{fang2024source,
  title={Source-free unsupervised domain adaptation: A survey},
  author={Fang, Yuqi and Yap, Pew-Thian and Lin, Weili and Zhu, Hongtu and Liu, Mingxia},
  journal={Neural Networks},
  volume={174},
  pages={106230},
  year={2024},
  publisher={Elsevier}
}

@article{li2024comprehensive,
  title={A comprehensive survey on source-free domain adaptation},
  author={Li, Jingjing and Yu, Zhiqi and Du, Zhekai and Zhu, Lei and Shen, Heng Tao},
  journal={IEEE Trans. on Pattern Analysis and Machine Intelligence},
  volume={46},
  number={8},
  pages={5743--5762},
  year={2024},
  publisher={IEEE}
}

@inproceedings{carion2020end,
  title={End-to-end object detection with transformers},
  author={Carion, Nicolas and Massa, Francisco and Synnaeve, Gabriel and Usunier, Nicolas and Kirillov, Alexander and Zagoruyko, Sergey},
  booktitle={European Conf. on computer vision},
  pages={213--229},
  year={2020},
  organization={Springer}
}

@article{mohammadi2026amar,
      title={AMAR: Lightweight Attention-Based Multi-User Activity Recognition from Wi-Fi CSI}, 
      author={Amirhossein Mohammadi and Hina Tabassum},
      year={2026},
      eprint={2605.20649},
      archivePrefix={arXiv},
      primaryClass={eess.SP},
      url={https://arxiv.org/abs/2605.20649}, 
}

@article{kuhn1955hungarian,
  title={The Hungarian method for the assignment problem},
  author={Kuhn, Harold W},
  journal={Naval research logistics quarterly},
  volume={2},
  number={1-2},
  pages={83--97},
  year={1955},
  publisher={Wiley Online Library}
}

@article{liang2023wiai,
  title={WiAi-ID: Wi-Fi-based domain adaptation for appearance-independent passive person identification},
  author={Liang, Ying and Wu, Wenjie and Li, Haobo and Han, Feng and Liu, Zhengqi and Xu, Pengfei and Lian, Xiaoli and Chen, Xiaojiang},
  journal={IEEE Internet of Things Journal},
  volume={11},
  number={1},
  pages={1012--1027},
  year={2023},
  publisher={IEEE}
}

@article{jiao2025robust,
  title={Robust Indoor Localization in Dynamic Environments: A Multi-source Unsupervised Domain Adaptation Framework},
  author={Jiao, Jiyu and Wang, Xiaojun and Han, Chengpei},
  journal={arXiv preprint arXiv:2502.07246},
  year={2025}
}

@article{tan2022commodity,
  title={Commodity {WiFi} sensing in ten years: Status, challenges, and opportunities},
  author={Tan, Sheng and Ren, Yili and Yang, Jie and Chen, Yingying},
  journal={IEEE Internet of Things Journal},
  volume={9},
  number={18},
  pages={17832--17843},
  year={2022},
  publisher={IEEE},
  doi={10.1109/JIOT.2022.3164569}
}

@article{wang2026survey,
  title={A survey on wi-fi sensing generalizability: Taxonomy, techniques, datasets, and future research prospects},
  author={Wang, Fei and Zhang, Tingting and Xi, Wei and Ding, Han and Wang, Ge and Zhang, Di and Cui, Yuanhao and Liu, Fan and Han, Jinsong and Xu, Jie and others},
  journal={IEEE Communications Surveys \& Tutorials},
  year={2026},
  publisher={IEEE}
}

@article{cao2019learning,
  title={Learning imbalanced datasets with label-distribution-aware margin loss},
  author={Cao, Kaidi and Wei, Colin and Gaidon, Adrien and Arechiga, Nikos and Ma, Tengyu},
  journal={Advances in neural information processing systems},
  volume={32},
  year={2019}
}


\begin{IEEEbiography}[{\includegraphics[width=1.25in,height=1.25in,clip,keepaspectratio]{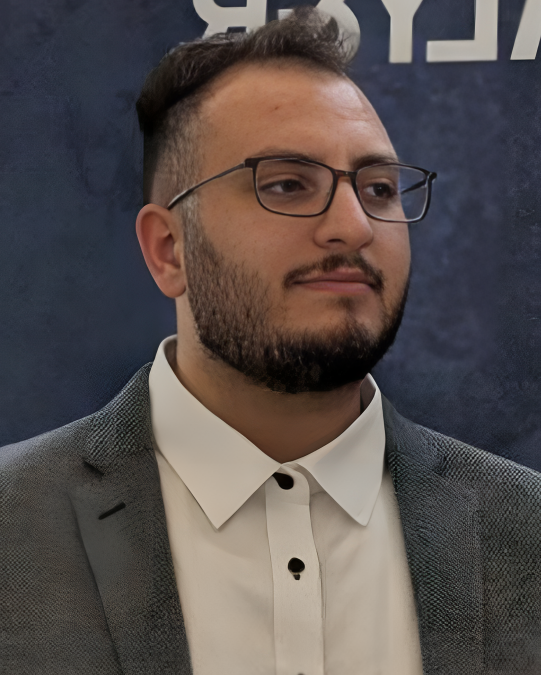}}]{Ahmed Radwan}
received the M.Sc. degree in computer science from York University, Toronto, ON, Canada, in 2026, under the supervision of Dr. H. Tabassum at the Next Generation Wireless Networks Lab, and the B.Sc. degree in computer science from King Abdulaziz University, Saudi Arabia, in 2024. From February to August 2024, he was a Visiting Research Student at King Abdullah University of Science and Technology (KAUST), under the supervision of Prof. Tareq Y. Al-Naffouri and Prof. Mohamed-Slim Alouini. He is currently an Applied Machine Learning Associate at the Vector Institute, Toronto, ON, Canada. His research interests span trustworthy and multimodal AI, with a focus on fairness, bias mitigation, generalization, and reliable deployment of AI systems across diverse real-world conditions, with applications to WiFi sensing, time-series analysis, and multimodal understanding.
\end{IEEEbiography}

\vskip -2\baselineskip plus -1fil

\begin{IEEEbiography}[{\includegraphics[width=1.25in,height=1.25in,clip,keepaspectratio]{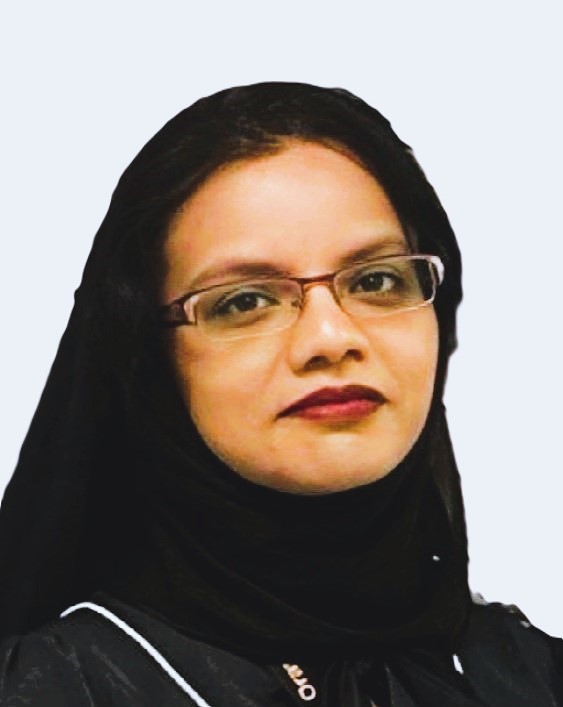}}]{Hina Tabassum} \,
(Senior Member, IEEE) \,  \, (M'12-SM'18)
received the Ph.D. degree from the King Abdullah University of Science and Technology (KAUST).  She is currently an Associate Professor with the Lassonde School of Engineering, York University, Canada, where she joined as an Assistant Professor in 2018. She is also appointed as a Visiting Faculty with the University of Toronto in 2024, and the York Research Chair of 5G/6G-enabled mobility and sensing applications in 2023, for five years. She is listed in the Stanford’s list of the World’s Top Two-Percent Researchers from 2021 to 2025. She has been selected as the IEEE ComSoc Distinguished Lecturer for the term 2025–2026. She has co-authored over 120 refereed articles in well-reputed IEEE journals, magazines, and conferences. Her current research interests include multiband 6G wireless communications and sensing networks, connected and autonomous systems, and AI-enabled network mobility and resource management solutions. She has earned numerous distinctions, including the N2Women Star in Networking and Communications (2025), Early Career Lassonde Innovation Award (2023), N2Women Rising Star in Networking and Communications (2022), multiple Exemplary Editor awards from IEEE journals, and appointment to the NSERC Discovery Grant Evaluation Group (2025–2028). She served as an Associate Editor for IEEE COMMUNICATIONS LETTERS from 2019 to 2023, IEEE OPEN JOURNAL OF THE COMMUNICATIONS SOCIETY from 2019 to 2023, and IEEE TRANSACTIONS ON GREEN COMMUNICATIONS AND NETWORKING from 2020 to 2023. She is also currently serving as an Area Editor for IEEE OPEN JOURNAL OF THE COMMUNICATIONS SOCIETY and an Associate Editor for IEEE TRANSACTIONS ON COMMUNICATIONS, IEEE TRANSACTIONS ON MOBILE COMPUTING, IEEE TRANSACTIONS ON WIRELESS COMMUNICATIONS, and IEEE COMMUNICATIONS SURVEYS AND TUTORIALS.
\end{IEEEbiography}

\vfill

\end{document}